\documentclass[yathesis,aps,preprintnumbers,11pt,amsmath,amssymb,nofootinbib]{revtex4}

\let\tmp\newinsert
\let\newinsert\newbox

\usepackage[utf8x]{inputenc} 
\usepackage{braket}
\usepackage{float}
\usepackage{graphics,epsfig}
\usepackage{bm}
\usepackage{slashed}
\usepackage{graphicx}
\usepackage{amsmath}
\usepackage{amsfonts}
\usepackage{amssymb}
\usepackage{color}
\usepackage{dcolumn}
\usepackage[
            pdfstartview=FitH,
            bookmarksnumbered=true,
            bookmarksopen=true,
            colorlinks,
            linkcolor=blue,
            anchorcolor=green,
            citecolor=red
            ]{hyperref}
\usepackage{enumerate}

\let\newinsert\tmp

\providecommand{\U}[1]{\protect\rule{.1in}{.1in}}

 \newcommand{\bq}{\begin{equation}}
 \newcommand{\eq}{\end{equation}}
 \newcommand{\bqn}{\begin{eqnarray}}
 \newcommand{\eqn}{\end{eqnarray}}
 \newcommand{\nb}{\nonumber}

\def\A0{A^{(0)}}

\begin{document}
\baselineskip=0.6 cm \title{Detecting sublunar-mass primordial black holes with the Earth-Moon binary system}

\author{Ya-Ling Li$^{1,2}$}
%\thanks{liyaling619@163.com}
\author{Guo-Qing Huang$^{1,2}$}
\thanks{huanggq@ncu.edu.cn}
\author{Zong-Qiang Huang$^{1,2}$}
\author{Fu-Wen Shu$^{1,2,3,4}$}
\thanks{shufuwen@ncu.edu.cn}

\affiliation{
	$^{1}$Department of Physics, Nanchang University, Nanchang, 330031, China\\
	$^{2}$Center for Relativistic Astrophysics and High Energy Physics, Nanchang University, Nanchang 330031, China\\
	$^{3}$GCAP-CASPER, Physics Department, Baylor University, Waco, Texas 76798-7316, USA\\
$^{4}$Center for Gravitation and Cosmology, Yangzhou University, Yangzhou, China}

\vspace*{0.2cm}

\begin{abstract}
\baselineskip=0.6 cm
\begin{center}
{\bf Abstract}
\end{center}
 In this paper we propose a new way to detect sublunar-mass primordial black holes (PBHs) by direct observations of the Earth-Moon binary system. Our method is based on treating PBH as a perturbation term,  by assuming that the PBH is far away from the Earth-Moon binary (far greater than $1$ AU) and the mass of the PBH is small (less than the lunar mass). This perturbation treatment allows us to develop a framework to calculate the orbits of a generic binary system such as the Earth-Moon binary system. Our numerical results show that the Earth-Moon distance is sensitive to the initial values of the system. In most cases,  the long-duration interactions between the PBH and the Earth-Moon system can induce lasting imprints on the Earth-Moon's orbit, and these imprints can accumulate over time, eventually giving rise to observable deviations which can be used to infer the properties of the PBH.
\end{abstract}

\maketitle
\vspace*{0.2cm}

\section{Introduction}
Astrophysical and cosmological observations provide convincing evidence that more than one quarter of the total energy density of
the Universe is  in the form of cold and weakly interacting matter, the dark matter (DM) \cite{Bertone:2018aa,Planck:2018vyg,Clowe:2006eq,Sofue:2000jx}. The experimental and theoretical searching for particle DM has last for several decades and a great deal of well-motivated particle DM candidates have been proposed. However, as the parameter space of particle DM models becomes tighter and tighter constrained by experiments \cite{Arcadi:2017kky,Roszkowski:2017nbc}, more and more nonparticle candidates are coming into view.  Among them primordial black holes (PBHs) as a candidate that could contribute a fraction of DM has attracted considerable attention in the past years. The research on PBHs could trace back to 1960s, which was initiated by Zel’dovich and Novikov \cite{osti_4471564}, and developed by Hawking and Carr \cite{Hawking:1971ei,Carr:1974nx,Carr:1975qj}.  It was generally believed that they could have been formed in the early Universe via the collapse of large density perturbations. Varies of mechanisms have been proposed to create such black holes in the past years \cite{hawking1989black,polnarev1991formation,yokoyama1995formation,garcia1996density,niemeyer1998near,clesse2015massive,garriga2016black,deng2017primordial,garcia2017primordial,gu2022primordial,cai2021one,cai2018primordial,chen2020dirac,fu2019primordial,lin2020primordial,yi2021primordial,yi2021primordial2,gao2021primordial,gao2021primordial2,fu2020primordial,meng2022one,zhou2020primordial,cai2020primordial,cai2021primordial,wu2021primordial,di2018primordial,chen2019primordial}. Most mechanisms predict that the mass of PBHs may exist in a wide range, from  $\sim 10^{14}$ g (below which they would have been evaporated completely via the Hawking radiation), to tens of thousands of solar masses.  Hence,  PBHs are a good candidate in explaining the origin of 
the black holes that cannot be formed in standard astrophysical processes, for instance, the intermediate mass black holes as recently observed by LIGO \cite{LIGOScientific:2020iuh,LIGOScientific:2020ufj}.

Various techniques have been developed in order to probe PBHs in a large range of masses in the past decades \cite{Katz:2018zrn,Munoz:2016tmg,Jung:2017flg,Jung:2019fcs,Laha:2018zav}. This process was further accelerated by the direct detection of gravitational waves \cite{LIGOScientific:2016aoc, LIGOScientific:2018jsj}. Up to now, there are numerous observational constraints on the
fraction of DM comprised of PBHs, see \cite{MACHO:2000qbb,carr1999dynamical,Afshordi:2003zb,Carr:2009jm,Carr:2016drx,Carr:2017jsz,Poulin:2017bwe,Carr:2020gox,Sasaki:2018dmp,Green:2020jor,EROS-2:2006ryy,Ali-Haimoud:2016mbv,Serpico:2020ehh,Hutsi:2019hlw,Wang:2016ana,Bird:2016dcv,Koushiappas:2017chw,Brandt:2016aco,Wang:2019kzb,Sasaki:2016jop,Ali-Haimoud:2017rtz,DeLuca:2020qqa,Jedamzik:2020ypm,Jedamzik:2020omx,Dasgupta:2019cae,chen2016constraint} for an incomplete list. However, there still exist a wide windows (masses from $10^{16}$g to $10^{21}$g) where PBHs can form whole or a significant fraction of the DM.  It is extremely difficult to detect these PBHs in a direct way, considering that they are only of atomic size. Despite all this, there are proposals that try to detect them in an indirect way. More specifically, these schemes consider that these PBHs are captured by neutron stars (NSs) or white dwarfs in dwarf galaxies \cite{Capela:2012jz,Capela:2013yf,Capela:2014ita}. It is generally believed that once a PBH is captured by a NS, the NS would be accreted onto the PBH such that NS gets destroyed in a much more short time than its normal lifetime. Hence, observations of NSs, in turn, will effectively impose 
constraints on the abundance of PBHs. This method, in a very recent, has been extended to the capture of PBHs by main sequence, Sun-like stars \cite{esser2022constraints}, which has the advantage,  as compared to NS and white dwarfs, that they can be observed in dwarf galaxies more frequently. However, these methods are model dependent, and, as mentioned before, are indirect. 

This motivates us to explore alternatives that can overcome these flaws. One possibility is to study the interaction between PBHs and astronomical binary systems, for instance, the Earth-Moon binary system, which has been studied for tens of centuries and is of the most accurate detections among all astronomical binary systems. The idea is simple. As PBHs pass nearby the Solar system, the long-duration interactions between PBHs and the Earth-Moon system could leave lasting imprints on the Earth-Moon's orbit. As the interactions continue, these imprints can accumulate over time, eventually giving rise to observable deviations which can be used to infer the properties of PBHs. Based on this idea, in this work we propose a new way to detect these sublunar-mass PBHs by direct observations of Earth-Moon system.  By treating PBH as a perturbative term, we also develop a new formalism for calculating the PBH-induced evolution of an astronomical binary system like the Earth-Moon system.

The remainder of this paper is organized as follows. In Sec. II we give a brief review on the the osculating orbital elements and the perturbed Kepler problem. In Sec. III we compute the evolution of the osculating orbital elements of a binary system due to the influence of the perturbing force produced by the PBH. In this section, we establish a model of the perturbed two-body problem to detect the PBH. In Sec. IV we show our numerical results for different initial conditions. We summarize our main results in the last section.

\section{FORMALISM}
In this section we would like to study the osculating orbital elements and the perturbed Kepler problem. We begin by briefly reviewing the fundamental properties of Keplerian orbits, and then introducing the description of Keplerian orbits in space relative to a reference frame and the equations of motion for the osculating orbital elements in the perturbed Kepler problem.

\subsection{Keplerian orbits}
The Keplerian orbits are determined by considering the motion of two bodies interacting only within the framework of the Newtonian gravity, and assuming that each body is taken to be spherically symmetric. The equation of motion is given by
    \begin{equation}
    \label{1}
     	\ddot{\mathbf{r}}=-\frac{\mu}{r^{3}} \mathbf{r} ,
    \end{equation}
where $\mu =GM_{tot}=G(m_{0} +m_{1}) $ and $r=\left | \mathbf{r}  \right |$, with $G$ being the universal gravitational constant, $M_{tot} $ being the total mass,  and $\mathbf{r}$ being the position vector from $m_{0} $ to $m_{1} $.

Eq. \eqref{1} indicates that the force is radial, which implies  $\mathbf{r} \times \ddot{\mathbf{r}}=\mathbf{d}  (\mathbf{r} \times \dot{\mathbf{r}} ) /dt=0$.
As a consequence, $\mathbf{H}\equiv\mathbf{r} \times \dot{\mathbf{r}}$ is a constant vector. The constancy of $\mathbf{H}$ has two direct results: One is that the motion is constrained in the orbital plane, a fixed plane which is normal to $\mathbf{H}$. The second is that the magnitude of  the vector can be written as
    \begin{equation}
    	\label{2}
    	H=r^{2}\dot{ \theta } , 
    \end{equation}
where $\theta $ is the position angle measured from some fixed line in the plane. 

With the help of Eq. \eqref{2}, one can get the shape of the orbit by solving the radial portion of Eq. \eqref{1}. The solution of Eq. \eqref{1} is given by \cite{brouwer2013methods}
    \begin{equation}
    	\label{3}
    	r=\frac{a\left(1-e^{2}\right)}{1+e\cos f }  ,
    \end{equation}
where $e$ is the eccentricity, and $f=\theta -\omega$ is the true anomaly, with $\omega$ being the argument of pericenter. Since $\omega$ is constant, the relation between the angle $f$ and time is given by Eq. \eqref{2},
    \begin{equation}
     	\frac{d f}{d t}=\frac{H}{r^{2}}  .
    \end{equation}

According to the Kepler's third law, we can get the relation between the period  and the semimajor axis,
    \begin{equation}
    	\frac{2 \pi}{P}=\sqrt{\frac{\mu}{a^{3}}} ,
    \end{equation}
where $P$ is the period, and $a$ is the semimajor axis of the elliptical orbit.

\subsection{Keplerian orbits in space}
\begin{figure}[h!]
	{\centering
		\includegraphics[width = 1\textwidth]{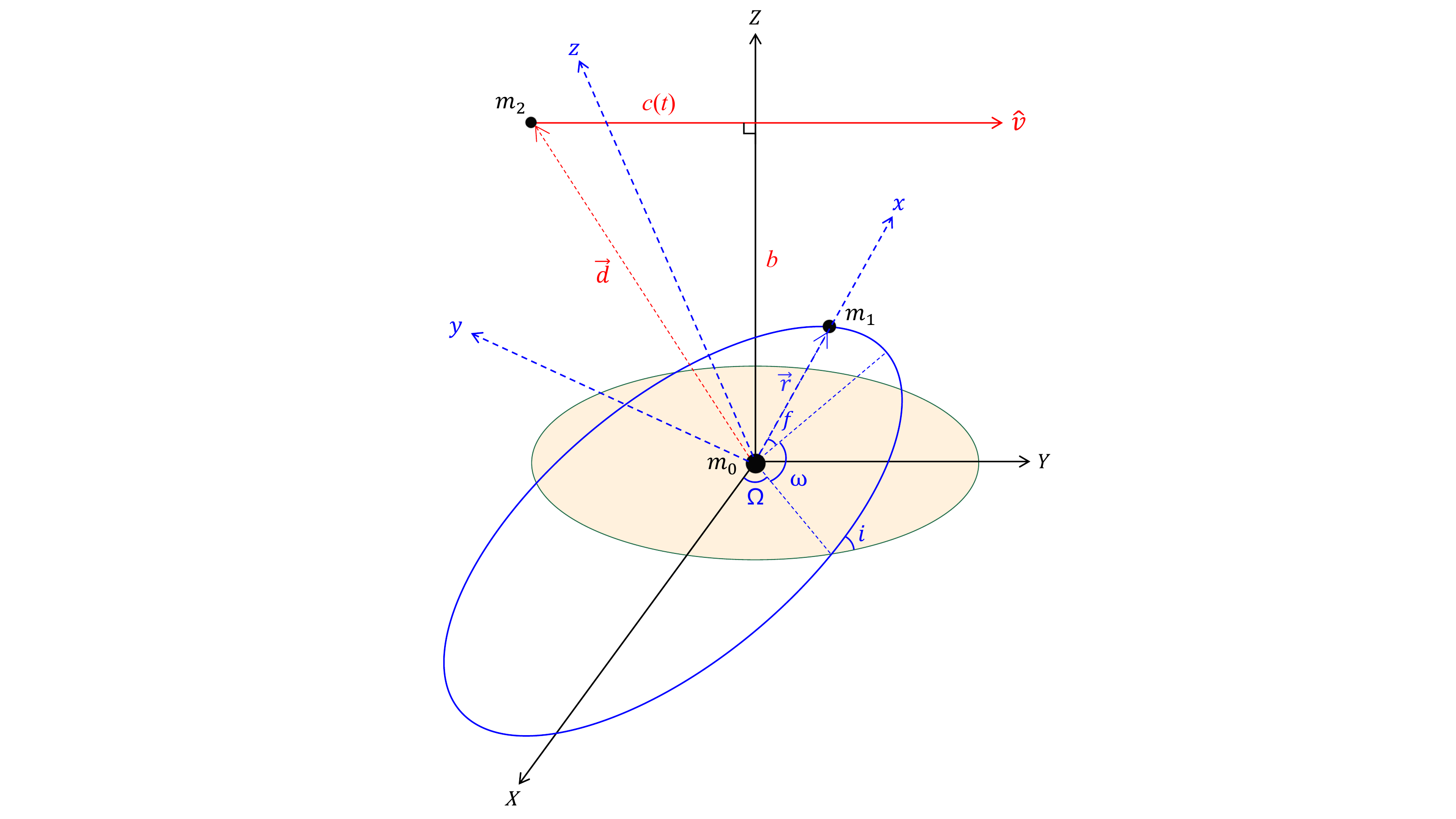}
	}
	\caption{Schematic picture of the model for detecting PBH with the Earth-Moon binary system. $m_{0}$, $m_{1}$, $m_{2}$ denote the Earth, the Moon, and the PBH, respectively. 
		The elliptical orbit of $m_{1}$ is described by the blue line in the figure. The trajectory of the $m_{2}$ is described by the solid red line, which always falls on the $Y$-$Z$ plane and parallels to the $Y$-axis. $\mathbf{d}=c(t)\mathbf{e} _{Y} +b\mathbf{e} _{Z} $ is the position vector from $m_{0} $ to $m_{2}$, with $c(t)$ and $b$ being the components of $\mathbf{d} $ on the $Y$-axes and $Z$-axes, respectively. $\mathbf{v} $ is the velocity vector of the $m_{2}$ and $\mathbf{r} $ is the position vector from $m_{0} $ to $m_{1} $.}
		\label{PBH-mode1}
\end{figure}

Let us first introduce a fundamental frame with coordinates ($X$, $Y$, $Z$) and an orbital frame with coordinates ($x$, $y$, $z$) as shown in Fig. \ref{PBH-mode1}. 
In the fundamental ($X$, $Y$, $Z$) frame, let us adopt the $X$-$Y$ plane as a reference plane, while the $Z$-axis as a reference direction. We also assign a constant vectorial basis $\mathbf{e} _{X}$, $\mathbf{e} _{Y}$ and $\mathbf{e} _{Z}$ to the fundamental frame. 
In the orbital ($x$, $y$, $z$) frame, the $x$-$y$ plane is the orbital plane of two-body motion, the $x$-direction is the radial direction of $m_{1}$ relative to $m_{0}$, and the $z$-direction is aligned with the angular-momentum vector. The orbital frame comes with a time-dependent basis vectors $\mathbf{e} _{x}$, $\mathbf{e} _{y}$ and $\mathbf{e} _{z}$.
The orientation of the elliptical orbit relative to the fundamental ($X$, $Y$, $Z$) frame is represented by the longitude of ascending node $\Omega$, the inclination $i$ and the argument of pericenter $\omega$.

With these definitions and conventions, we can go from the orbital ($x$, $y$, $z$) frame to the fundamental ($X$, $Y$, $Z$) frame by performing three consecutive Euler rotations,
\bqn
\label{6}\mathbf{e} _{X}&=&\left[\cos \Omega  \cos (\omega+f)-\cos i  \sin \Omega  \sin (\omega+f)\right]  \mathbf{e}_{x}
\nb\\&&+[-\cos \Omega  \sin (\omega+f)-\cos i  \sin \Omega \cos (\omega+f)]  \mathbf{e}_{y}	
\nb\\&&+\sin i  \sin \Omega\mathbf{e}_{z}, \\
\label{7}\mathbf{e} _{Y}&=&[\sin \Omega  \cos (\omega+f)+\cos i  \cos \Omega  \sin (\omega+f)]  \mathbf{e}_{x}
\nb\\&&+[-\sin \Omega  \sin (\omega+f)+\cos i  \cos \Omega  \cos (\omega+f)] \mathbf{e}_{y}
\nb\\&&-\sin i  \cos \Omega  \mathbf{e}_{z}, \\
\mathbf{e} _{Z}&=&\sin i  \sin (\omega+f)  \mathbf{e}_{x}+\sin i  \cos (\omega+f) \mathbf{e}_{y}+\cos i \mathbf{e}_{z}.
\eqn

Therefore, the elliptical orbit in space can be described in terms of six orbital elements: $a$, $e$, $i$, $\Omega $, $\omega $, $M$, which are called semi-major axis, eccentricity, inclination, longitude of ascending node, argument of pericenter, mean anomaly, respectively \cite{burns1976elementary}.

\subsection{Osculating Orbital Elements and the Perturbed Kepler Problem}
Let us return to the two-body problem, but now suppose that the binary is subjected to some small perturbing force \cite{hui2013binary}. The equation of motion is
\begin{equation}
	\ddot{\mathbf{r}}=-\frac{\mu}{r^{3}} \mathbf{r}+\mathbf{F}  ,
\end{equation}
where $\mathbf{F}$ is the perturbing acceleration. We decompose $\mathbf{F}$ as
\begin{equation}
	\mathbf{F}=R \boldsymbol{e}_{x}+T \boldsymbol{e}_{y}+N \boldsymbol{e}_{z}  ,
\end{equation}
in terms of components $R$, $T$ and $N$. Under the action of perturbing acceleration, the binary will deviate
from its Keplerian ellipse, causing its orbital elements to vary. We thus treat ($a$, $e$, $i$, $\Omega $, $\omega $, $M$) as functions of time, called the osculating orbital elements.

Following Refs. \cite{poisson2014gravity, murray1999solar}, we write the final equations for the osculating orbital elements,
\bqn
\label{11} \frac{d a}{d t}&=&\frac{2}{n \beta}[\mathrm{R} e \sin f+T(1+e \cos f)],\\
\label{12} \frac{d e}{d t}&=&\frac{\beta}{n a}\left[\mathrm{R} \sin f+T\left(\cos f+\frac{\cos f+e}{1+e  \cos f}\right)\right],\\
\frac{d i}{d t}&=&N \frac{\beta}{n a} \frac{\cos (\omega+f)}{1+e  \cos f},\\
\frac{d \Omega}{d t}&=&N \frac{\beta}{n a} \frac{\sin (\omega+f)}{1+e  \cos f} \csc i,\\
\frac{d \omega}{d t}&=&\frac{\beta}{n a e}[-\mathrm{R} \cos f+T(1+\gamma) \sin f]-\cos i \frac{d \Omega}{d t},\\
\frac{d M}{d t}&=&n+\frac{\beta^{2}}{n a e}[\mathrm{R}(\cos f-2 \gamma e)-T(1+\gamma) \sin f],
\eqn
with
\begin{equation}
	\label{17}
	\frac{d f}{d t}=\frac{n}{\beta^{3}}(1+e \cos f)^{2}+\frac{\beta }{n a e}\left[R \cos f-T \left(1+\gamma\right)\sin f\right] ,
\end{equation}
where $\beta=\sqrt{1-e^{2}}$, $\gamma=\frac{1}{1+e\cos f}$ and $n=\frac{2 \pi}{P}$.

\section{Detecting PBH with binary system: the model}
 In this section we turn to calculate the evolution of the osculating orbital elements of a binary system due to the influence of the perturbing force produced by a PBH. We first introduce the model for detecting PBH with the Earth-Moon binary system, before obtain the analytical expression of the perturbing force, and the evolution of the osculating orbital elements with time under different initial conditions by numerical calculation. We also discuss the dependence of the Earth-Moon distance variation $\Delta r$ on $\varphi$, $i$, $\Omega$, $\omega$, $b$, and $m_{2}$ under different initial conditions.
 
 Let us assume that the mass of the PBH is very small compared to the Earth and the Moon, and  the distance of the PBH is very far away from the Earth-Moon binary system, then we can establish a model of the perturbed two-body problem for detecting the PBH, as sketched in Fig. \ref{PBH-mode1}. 
In this model, the Earth is placed as the origin of the frame 
\footnote{We also assume that the center of mass of the binary is on the Earth considering the small mass of the Moon compared to the Earth.}, the Moon performs the Kepler motion relative to the Earth, and the force of the PBH on the Moon is regarded as a perturbing one.

The perturbing acceleration is then given by
\begin{equation}
	\mathbf{F}  =\frac{G m_{2}(\mathbf{d} -\mathbf{r} )}{|\mathbf{d} -\mathbf{r} |^{3}},
\end{equation}
where $\mathbf{r} =r \mathbf{e} _{x}$ and $\mathbf{d} =c(t)  \mathbf{e} _{Y}+b  \mathbf{e} _{Z}$. Suppose that the observation starts from the moment $c(0)=-100b$, then $c(t)=vt-100b$ with $v$=350 km$\cdot$s$^{-1}$, which is the typical velocity for halo dark matter relative to the solar system \cite{carr1999dynamical}. Since $d(t)=\sqrt{b^2+c(t)^2}\gg r$, $\mathbf{F}$ can be approximated as
\begin{equation}\label{Fapp}
\mathbf{F}  \approx \frac{G m_{2}}{d(t)^{3}}\left[c(t) \mathbf{e} _{Y}+b \mathbf{e} _{Z}-r \mathbf{e} _{x}\right].
\end{equation}

Substituting Eqs. \eqref{6} and \eqref{7} into  Eq. \eqref{Fapp}, we have
\begin{eqnarray}
	\mathbf{F}&\approx&\frac{G m_{2}}{d(t)^{3}}
 \left\{
	\Big(c(t)[\sin \Omega \cos (\omega+f)+\cos i \cos \Omega \sin (\omega+f)]+b \sin i \sin (\omega+f)-r\Big) \mathbf{e} _{x} \right. \nonumber\\
	&&\left.+\Big(c(t)[-\sin \Omega \sin (\omega+f)+\cos i \cos \Omega \cos (\omega+f)]+b \sin i \cos (\omega+f)\Big) \mathbf{e} _{y}\right. \nonumber \\
	&&\left.+\left[ c(t)(-\sin i \cos \Omega)+b \cos i\right]  \mathbf{e} _{z}
\right\}.
\end{eqnarray}
Since $\mathbf{F}=R \mathbf{e}_{x}+T \mathbf{e}_{y}+N \mathbf{e}_{z}$, the components of the perturbing acceleration in the basis vectors ($\mathbf{e} _{x}$, $\mathbf{e} _{y}$, $\mathbf{e} _{z}$) are
\bqn
\nb R&=&\frac{G m_{2}}{d(t)^3}\{c(t)[\sin \Omega \cos (\omega+f)+\cos i \cos \Omega \sin (\omega+f)]+b \sin i \sin (\omega+f)-r\},\\
\nb \\
\nb T&=&\frac{G m_{2}}{d(t)^3}\{c(t)[-\sin \Omega \sin (\omega+f)+\cos i \cos \Omega \cos (\omega+f)]+b \sin i \cos (\omega+f)\},\\
\nb \\
 \label{21} N&=&\frac{G m_{2}}{d(t)^3}[c(t)(-\sin i \cos \Omega)+b \cos i],
\eqn 
where $i$, $\Omega$ and $\omega$ are parameters describing elliptical orbit of the Moon. We can change the trajectory of the PBH relative to the Earth-Moon binary system by changing the initial values of $i$, $\Omega$ and $\omega$.  
Substituting Eq. \eqref{21} into Eqs. \eqref{11}-\eqref{17}, we can get the variation of ($a$, $e$, $i$, $\Omega $, $\omega $, $M$) with time.

Since the orbit under the perturbing force is tangent to the Keplerian ellipse at each moment, and the position and velocity of the particle in the real orbit are consistent with the corresponding point in the osculating orbit at that moment, so the relationship between $a$, $e$ and $r$ still satisfies the two-body motion relationship Eq. \eqref{3}. Differentiating Eq. \eqref{3}, we can get the variation of $r$ with time.

\section{Numerical results}
\subsection{When the initial position is coplanar}
\begin{figure}[h!]
	{\centering
		\includegraphics[width = 1\textwidth]{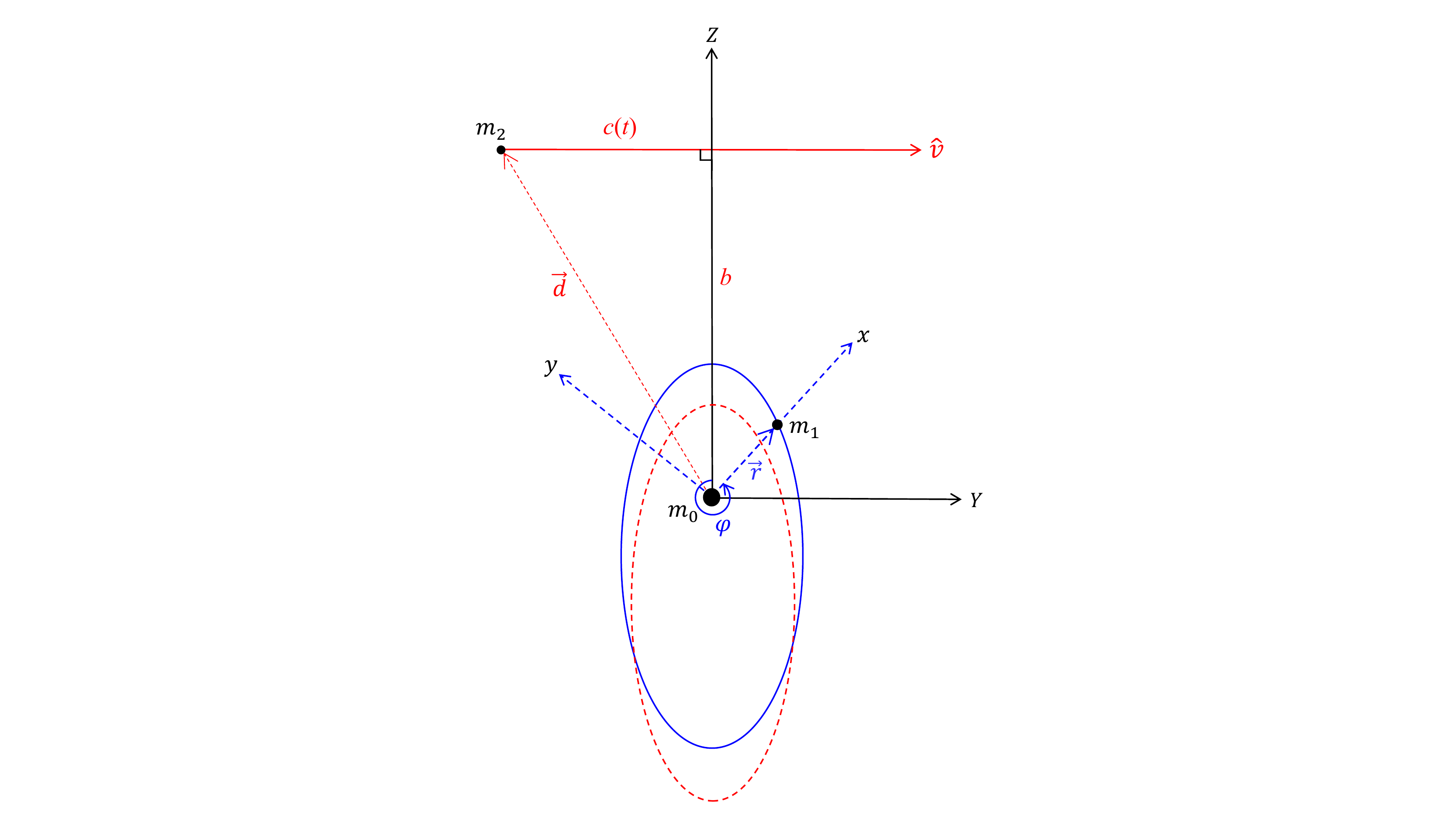}
	}
	\caption{Schematic diagram when the trajectory of the PBH is coplanar to the elliptical orbit of the Moon. The elliptical orbit of the Moon is depicted by the blue solid line, when $c(t)=-100b$. The elliptical orbit of the Moon is depicted by the red dotted line, when $c(t)=0$.}
	\label{PBH-mode2}
\end{figure}

We first discuss the influence of the perturbing force of the PBH on the Earth-Moon binary system, when the trajectory of the PBH is coplanar to the elliptical orbit of the Moon, as shown in Fig. \ref{PBH-mode2}. 
Throughout the calculations of this subsection, we set the following initial condition: $i=\frac{\pi }{2}$, $\Omega =\frac{\pi }{2}$, $\omega =\frac{\pi }{2}$, $b=10$AU and $m_{2} =10^{20}$kg. 

\begin{figure*}[htbp]
	\centering
	\includegraphics[scale=0.3]{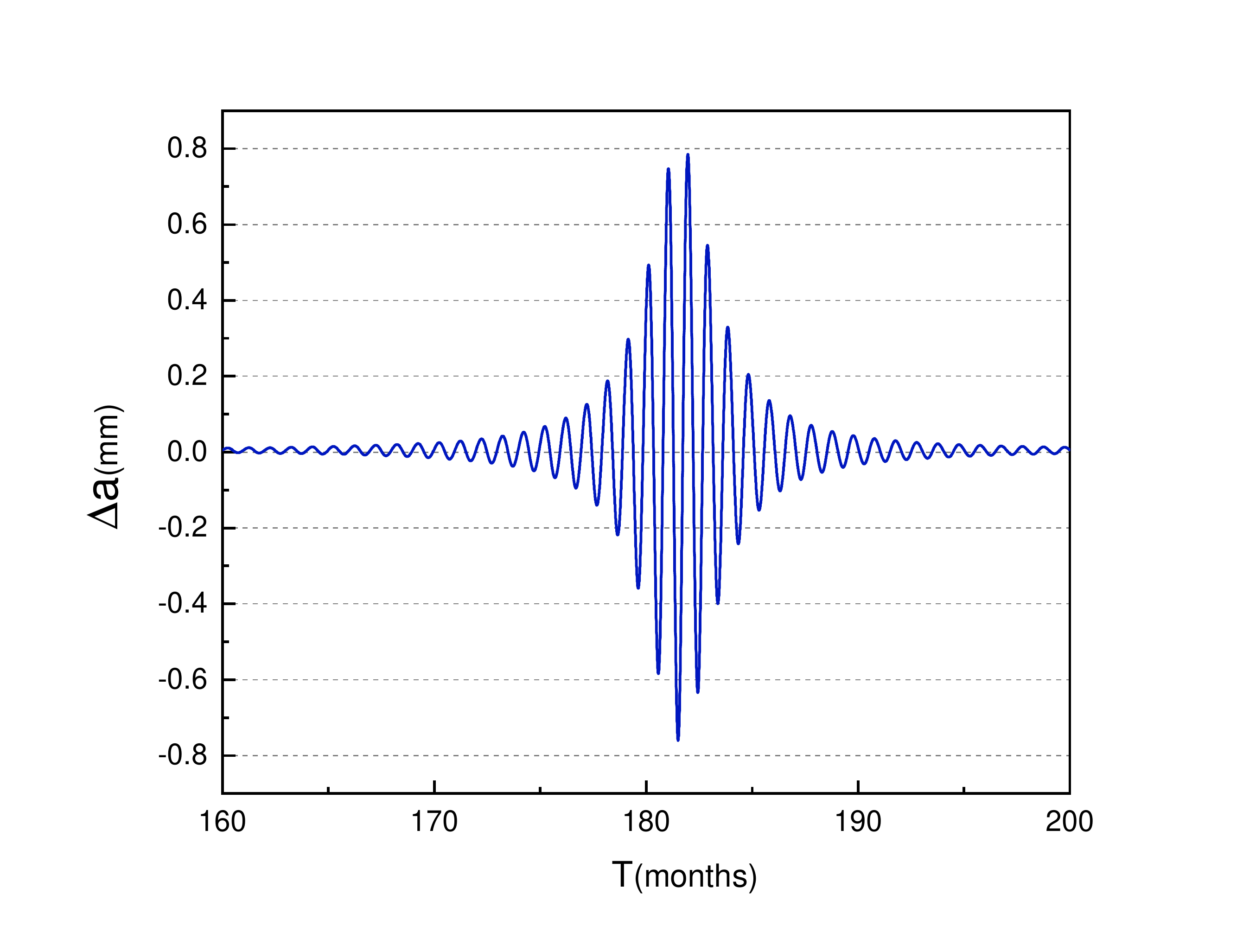}
	\includegraphics[scale=0.3]{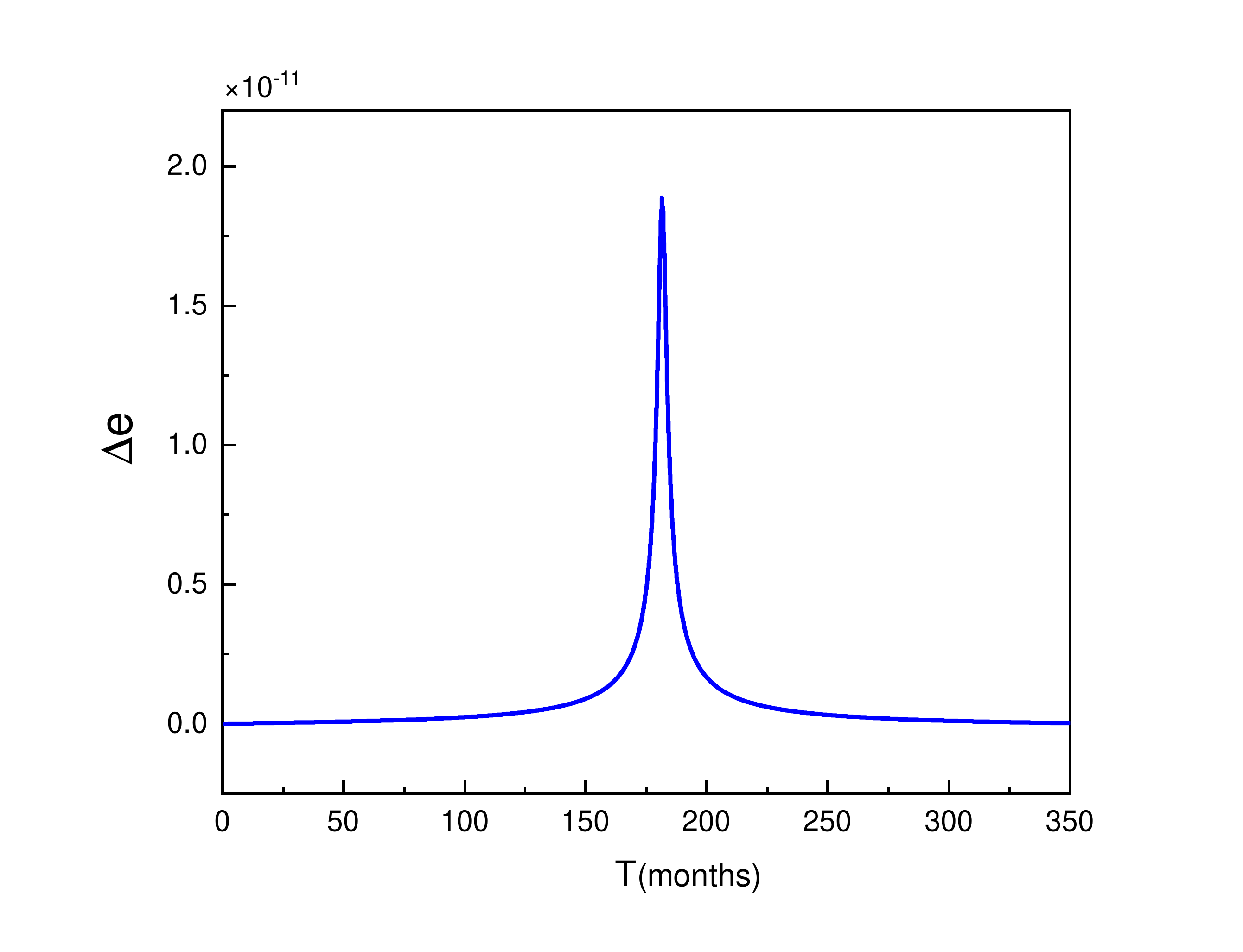}
	\includegraphics[scale=0.3]{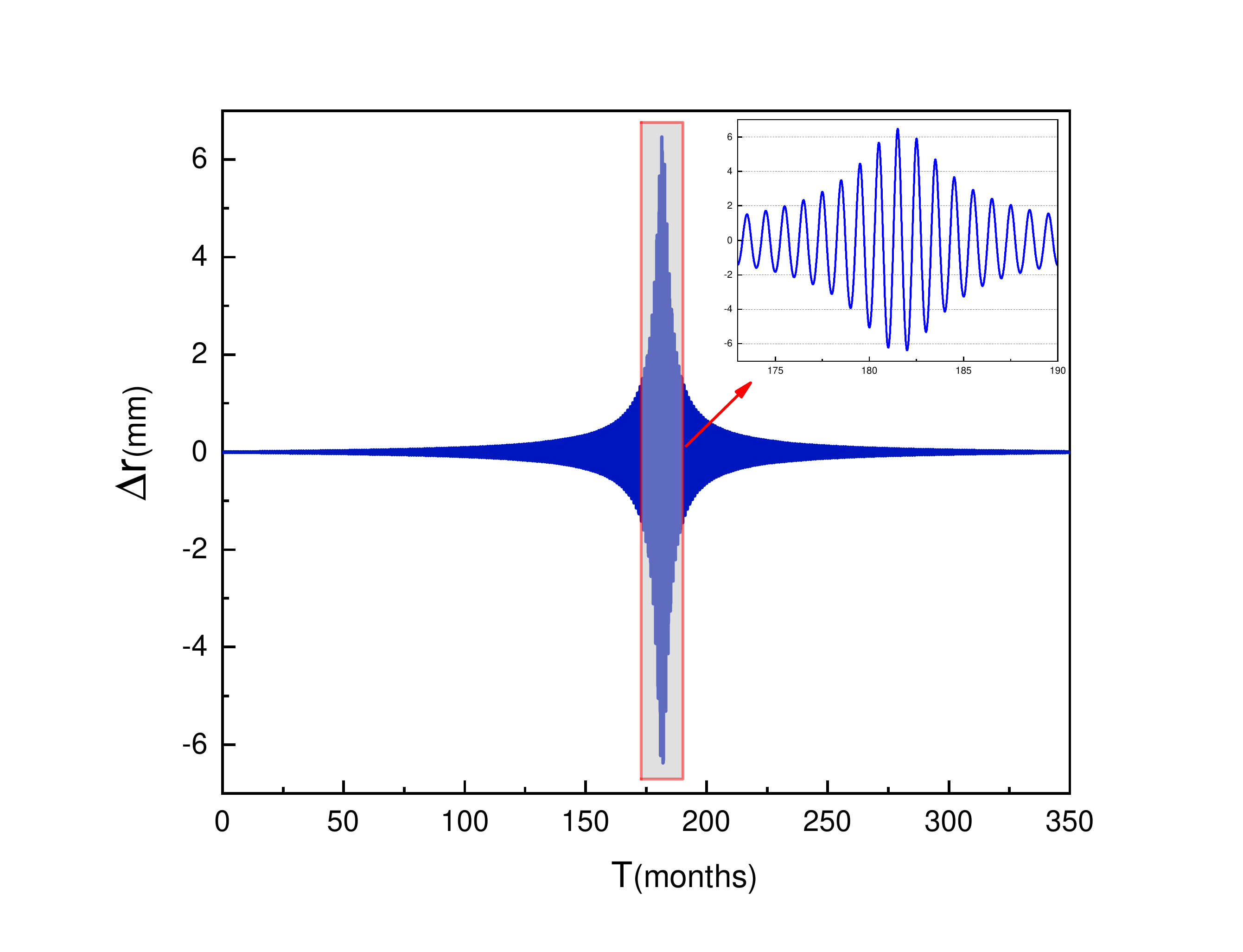}
	\caption{The evolution of $\Delta a$, $\Delta e$, $\Delta r$ with time. }
	\label{PBH-3}
\end{figure*}

Substituting the initial condition into Eq. \eqref{21}, the components of the perturbing acceleration are given by
\bqn
R&=&\frac{G m_{2}}{d(t)^3}[c(t)(-\sin f)+b \cos f-r],\\
T&=&\frac{G m_{2}}{d(t)^3}[c(t)(-\cos f)-b \sin f],\\
N&=&0.
\eqn 
Plugging them into Eqs. \eqref{11} and \eqref{12}, the rate of change of $a$ and $e$ then become
\bqn
\label{25} \frac{d a}{d t}&=&-\frac{2}{n \beta} \frac{G m_{2}}{d(t)^3}[c(t)(e+\cos f)+(r e+b) \sin f],\\
\label{26} \nb \frac{d e}{d t}&=&-\frac{\beta}{n a} \frac{G m_{2}}{d(t)^3} \frac{c(t)[1+2 e \cos f+\cos ^{2}  f]+[r+be+(r e+b) \cos f] \sin f}{1+e \cos f}.\\
\eqn
Combined with Eq. \eqref{3}, evolutions of $\Delta a$, $\Delta e$, $\Delta r$ with time are obtained by numerical calculations. The results are shown in Fig. \ref{PBH-3}. 
From these figures, we can see that both $\Delta a$ and $\Delta r$ are oscillating quasiperiodically over time, and their amplitudes first increase and then decrease. $\Delta e$, in contrast, increases and then decreases over time monotonously. It is worth noting that all of them ($\Delta a$, $\Delta e$ and $\Delta r$) reach to zero in the end, which means that the perturbing force of PBH has no long-term effects on the elliptical orbit of the Moon.
 The significance here can be understood by analyzing
  Eqs. \eqref{3}, \eqref{25} and \eqref{26}. 
First, since the term $\sin{f}$ is a periodic function with time, its time integral over the total process is equal to zero.
Then, since the trajectory of the PBH is symmetrical about the $Z$-axis, the time integral of the term $c(t)=vt-100b$ over the total process is also equal to zero.
Therefore, the time integral of Eq. \eqref{25} over the total process will be equal to zero, which means that the semi-major axis of the elliptical orbit does not change in the end.
In the same way, we can also understand both $\Delta e$ and $\Delta r$ are equal to zero in the end, by analyzing the Eq. \eqref{26} and Eq. \eqref{3}.

\begin{figure*}[htbp]
	\centering
	\includegraphics[scale=0.3]{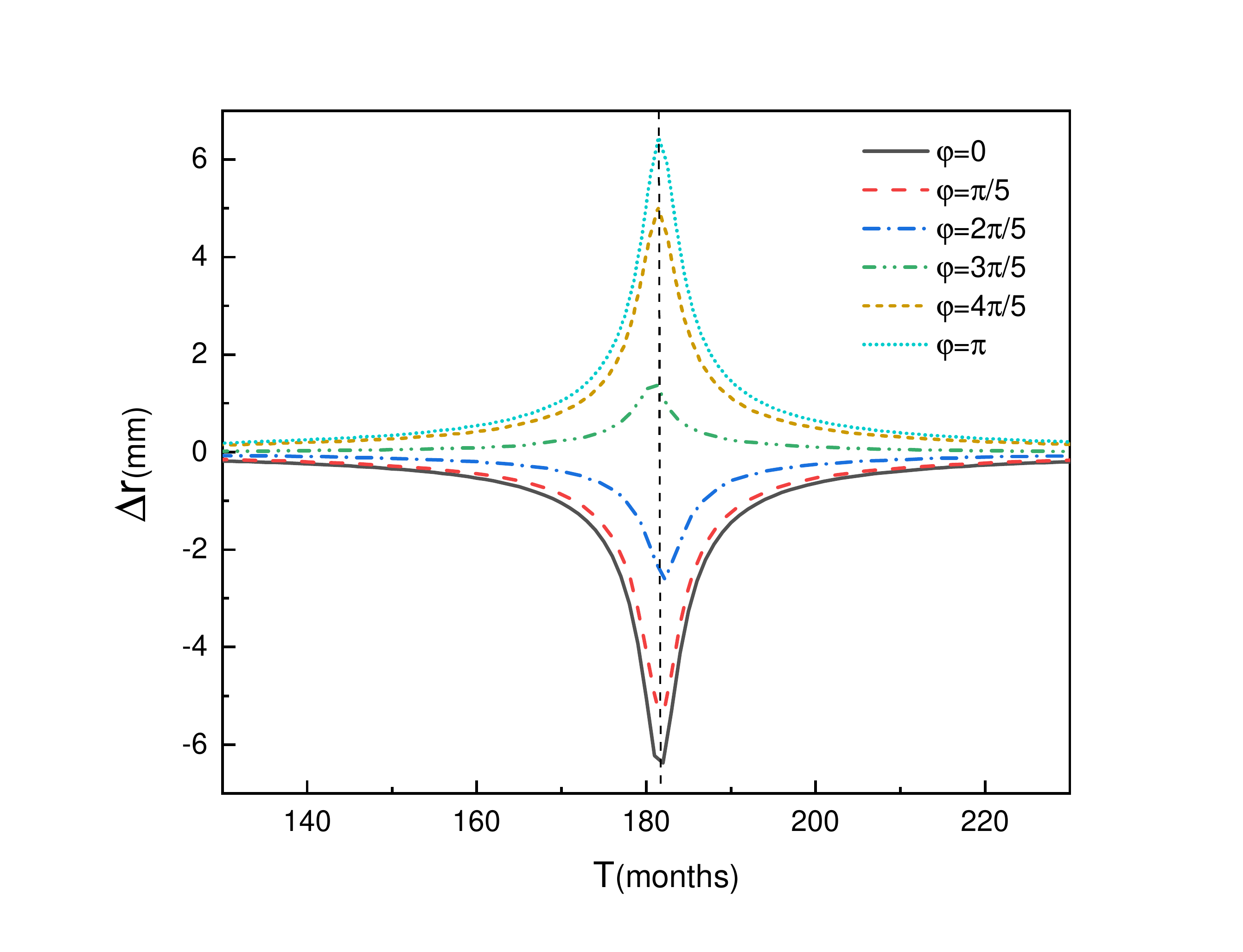}
	\includegraphics[scale=0.3]{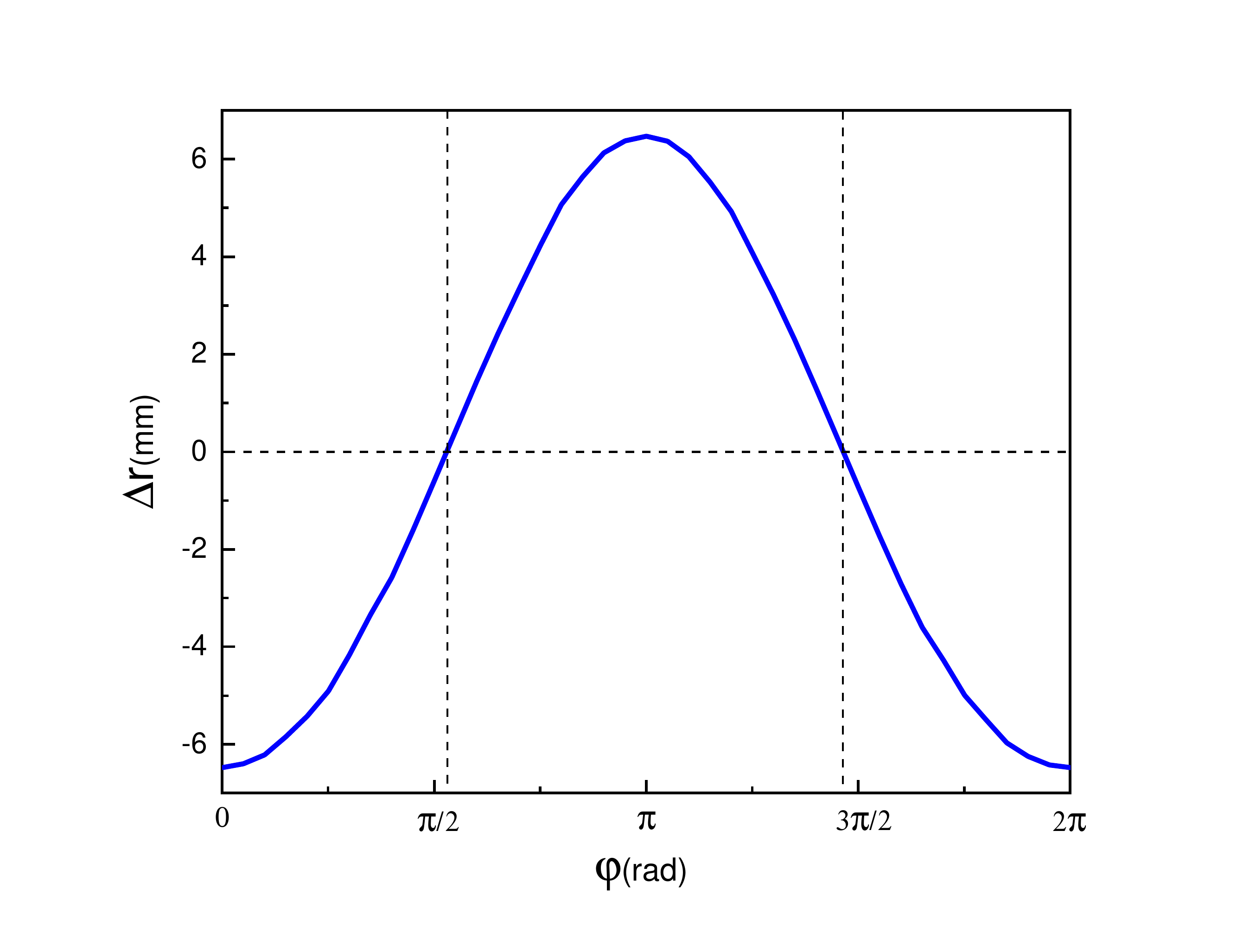}
	\caption{Left panel: the evolution of $\Delta r$ with time under different fixed observation angles $\varphi$. Right panel: the peak value of $\Delta r(T)$ (i.e., the values of $\Delta r$ at the dotted line in the figure on the left panel) as a function of $\varphi$, which corresponds to the dependence of $\Delta r$ on $\varphi$ when $c(t)=0$.}
	\label{PBH-4}
\end{figure*}

In order to quantitatively study the change of the distance between the Earth and the Moon under the perturbing force of PBH, we calculate the evolution of $\Delta r$ with time under a fixed observation angle $\varphi $. 
The observation angle $\varphi $ is defined as
\begin{equation}
f=2m\pi +\varphi,
\end{equation}
where $m=0,1,2...$ and $\varphi \in [ 0,2\pi )$.
For instance, $\varphi=0$ and $\varphi=\pi$ corresponds to the moments when the Moon is at perigee and apogee, respectively, which are called ``normal points''. In the case of $\varphi=0$, the evolution of the distance between the Earth and the Moon corresponds to make measurements when the Moon is at perigee. The rate of change of $a$ and $e$ in this case are then given by
\bqn
\label{28} \frac{d a}{d t}&=&-\frac{2}{ n \beta} \frac{G m_{2}}{d(t)^3}(1+e)c(t),\\
\label{29} \frac{d e}{d t}&=&- \frac{2\beta}{n a} \frac{G m_{2}}{d(t)^3}c(t).
\eqn
Finally, the evolution of $\Delta r$ with time can be obtained by numerical calculating Eqs. \eqref{3}, \eqref{28} and \eqref{29}. The results are presented as the black solid line in Fig. \ref{PBH-4}.
From the black solid line with $\varphi=0$, we can see that $\left | \Delta r \right | $ increases first and then decreases over time monotonously, and reaches a maximum at a certain moment.
The significance here can be understood by analyzing the perturbing force of PBH on the Moon.
  When the Moon is at perigee with $\varphi=0$ in Fig. \ref{PBH-mode2}, the direction of the Moon's velocity is along the negative direction of the $Y$-axis and the perturbing force of PBH on the Moon can be decomposed into the component $R\mathbf{e} _{Y}$ along $Y$-axis direction and the component $T\mathbf{e} _{Z}$ along the $Z$-axis direction.
 The component $R\mathbf{e} _{Y}$ is consistent with the direction of the Moon's velocity when $c (t)< 0$, which causes the Moon's velocity  to increase. The greater the velocity, the smaller the orbital radius, according to the law of universal gravitation. The above means that $\left | \Delta r \right | $ increases with time when $c (t)< 0$.
  Similarly, it can be analyzed that
 the component $R\mathbf{e} _{Y}$ is opposite to the direction of the Moon's velocity when $c (t)> 0$, which causes $\left | \Delta r \right | $ to decrease with time.
 The component $T\mathbf{e} _{Z}$ is perpendicular to the direction of the Moon's velocity throughout the process, which does not change the magnitude of the Moon's velocity.
 Therefore, $\left | \Delta r \right | $ increases and then decreases over time monotonously throughout the process, and reaches the maximum when $c(t)=0$. 

 In order to quantitatively analyze the relationship between $\Delta r$ and $\varphi$, let us plot the evolution of $\Delta r$ with time at different observation angles $\varphi$ as shown in the left panel of Fig. \ref{PBH-4}, and the dependence of the maximum of $\Delta r$ on $\varphi$ as shown in the right panel of Fig. \ref{PBH-4}. 
From the right panel of Fig. \ref{PBH-4}, we can see that $\Delta r$ acts like a trigonometric function of $\varphi$. 
 The values of $\Delta r  $  is equal to zero when $\varphi$ tends to $\frac{\pi }{2}$ and $\frac{3\pi }{2}$, and $\left | \Delta r \right | $ reaches its maximum value ($\approx 6.46$mm) when  $\varphi$ is equal to $0$ or $\pi $. 
This means that the greatest change in the distance between the Earth and the Moon can be observed by the ``normal point" measurement. Since the current detection accuracy of the Earth-Moon distance can reach the millimeter level \cite{murphy2008apache,battat2009apache,murphy2012apollo,murphy2013lunar}, we can accurately detect PBH with the Earth-Moon binary system, when the system reaches certain initial conditions.
 
Recalling the change curve of $e$ with time in Fig. \ref{PBH-3} and the above analysis, we can roughly speculate that the elliptical orbit of the Moon first moved down as a whole and then moved up back to its initial position over time. And the offset and eccentricity of the elliptical orbit reach the maximum when $c(t)=0$. 
In the end, we can roughly represent the elliptical orbit of the Moon with the red dotted line in the Fig. \ref{PBH-mode2}.

\begin{figure*}[htbp]
	\centering
    \includegraphics[scale=0.3]{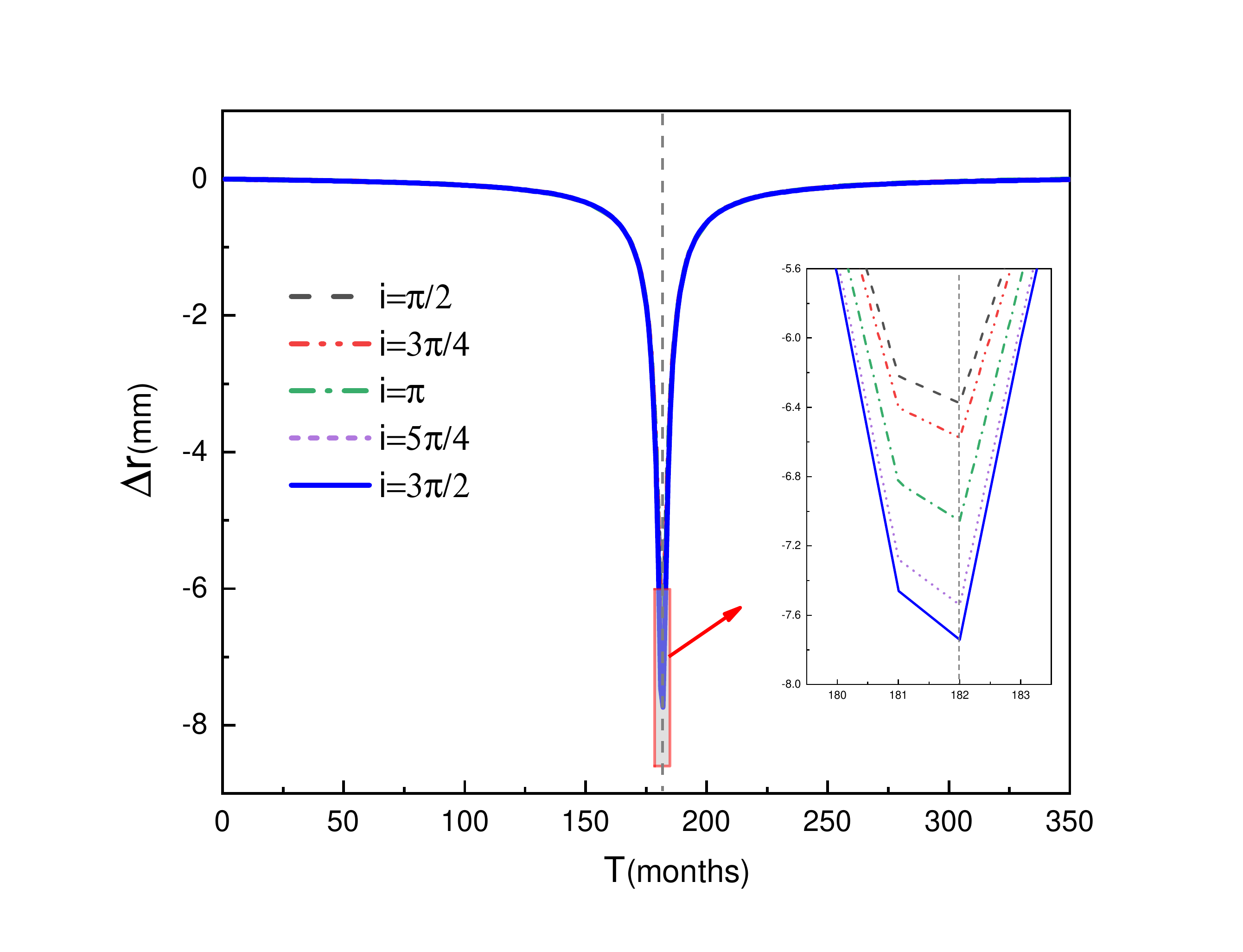}
    \includegraphics[scale=0.3]{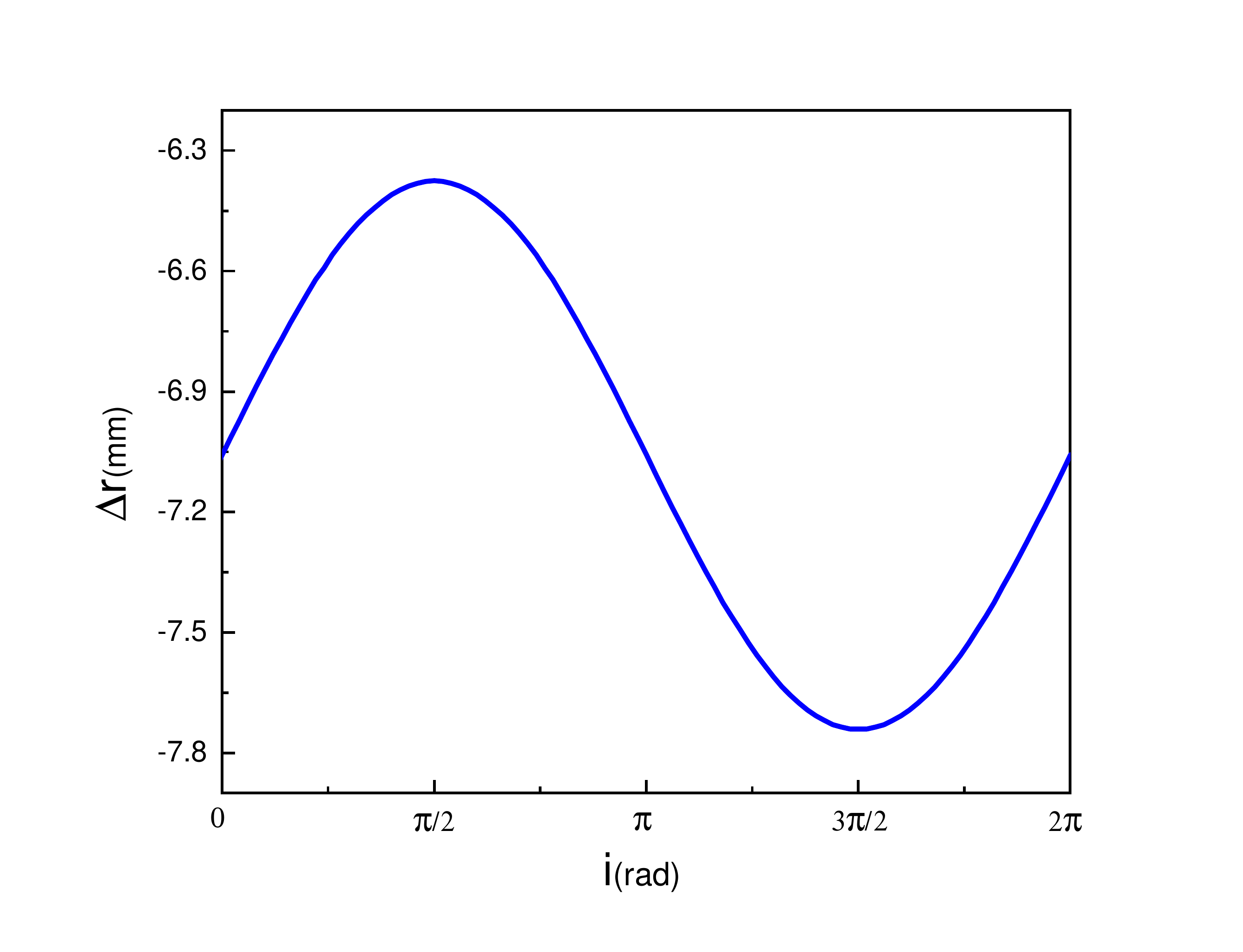}
	\caption{Left panel: the evolution of $\Delta r$ over time when $\varphi=0$ under different $i$ and other parameters fixed. Right panel:
	the peak value of $\Delta r(T)$ (i.e., the values of $\Delta r$ at the dotted line in the figure on the left panel) as a function of $i$, which corresponds to the dependence of $\Delta r$ on $i$ when $c(t)=0$	.}
	\label{PBH-5}
\end{figure*}
\begin{figure*}[htbp]
	\centering
	\includegraphics[scale=0.3]{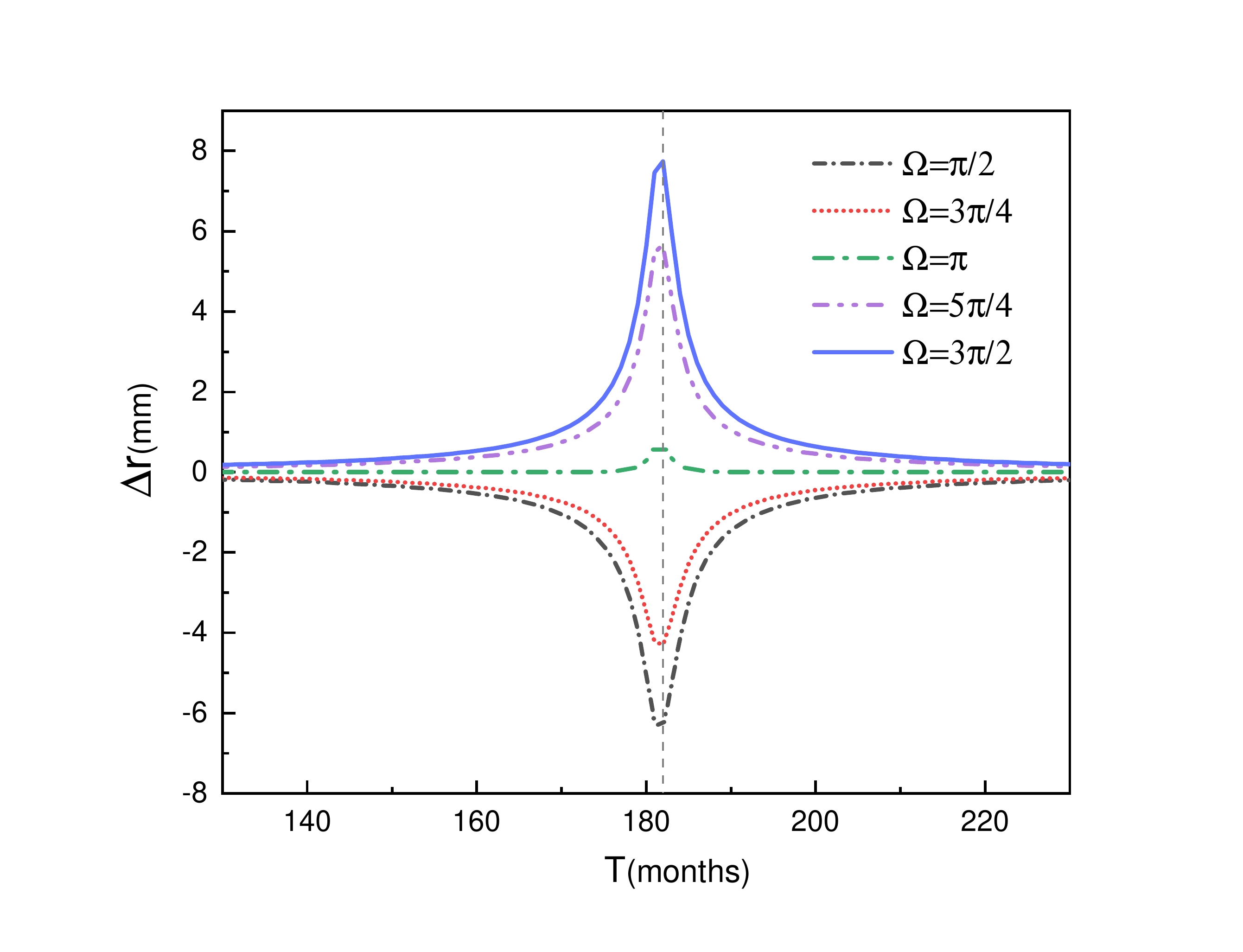}
	\includegraphics[scale=0.3]{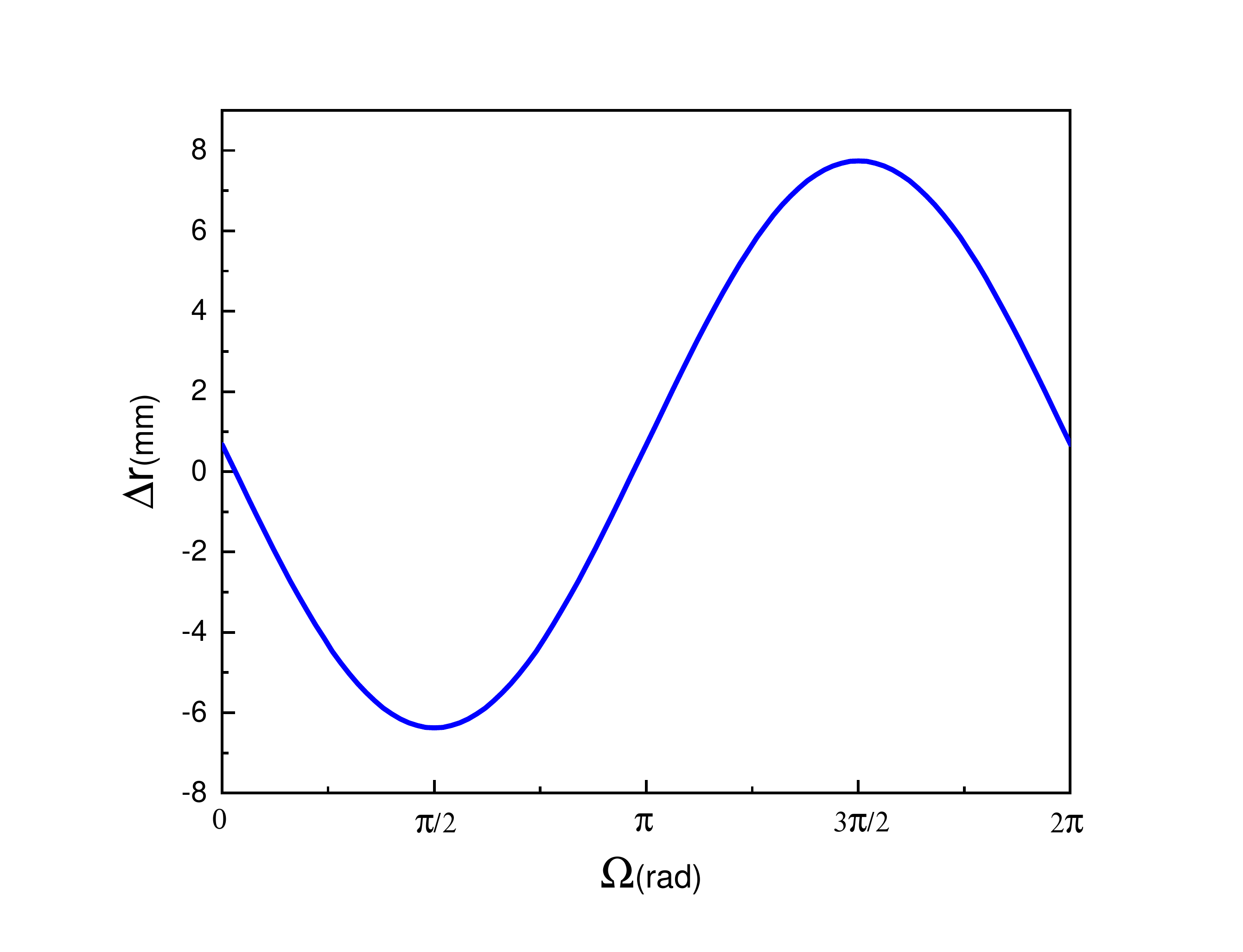}
	\caption{Left panel: the evolution of $\Delta r$ with time when $\varphi=0$ under different $\Omega$ and other parameters fixed. Right panel: 
		the peak value of $\Delta r(T)$ (i.e., the values of $\Delta r$ at the dotted line in the figure on the left panel) as a function of $\Omega$, which corresponds to the dependence of $\Delta r$ on $\Omega$ when $c(t)=0$.}
	\label{PBH-6}
\end{figure*}
\begin{figure*}[htbp]
	\centering
	\includegraphics[scale=0.3]{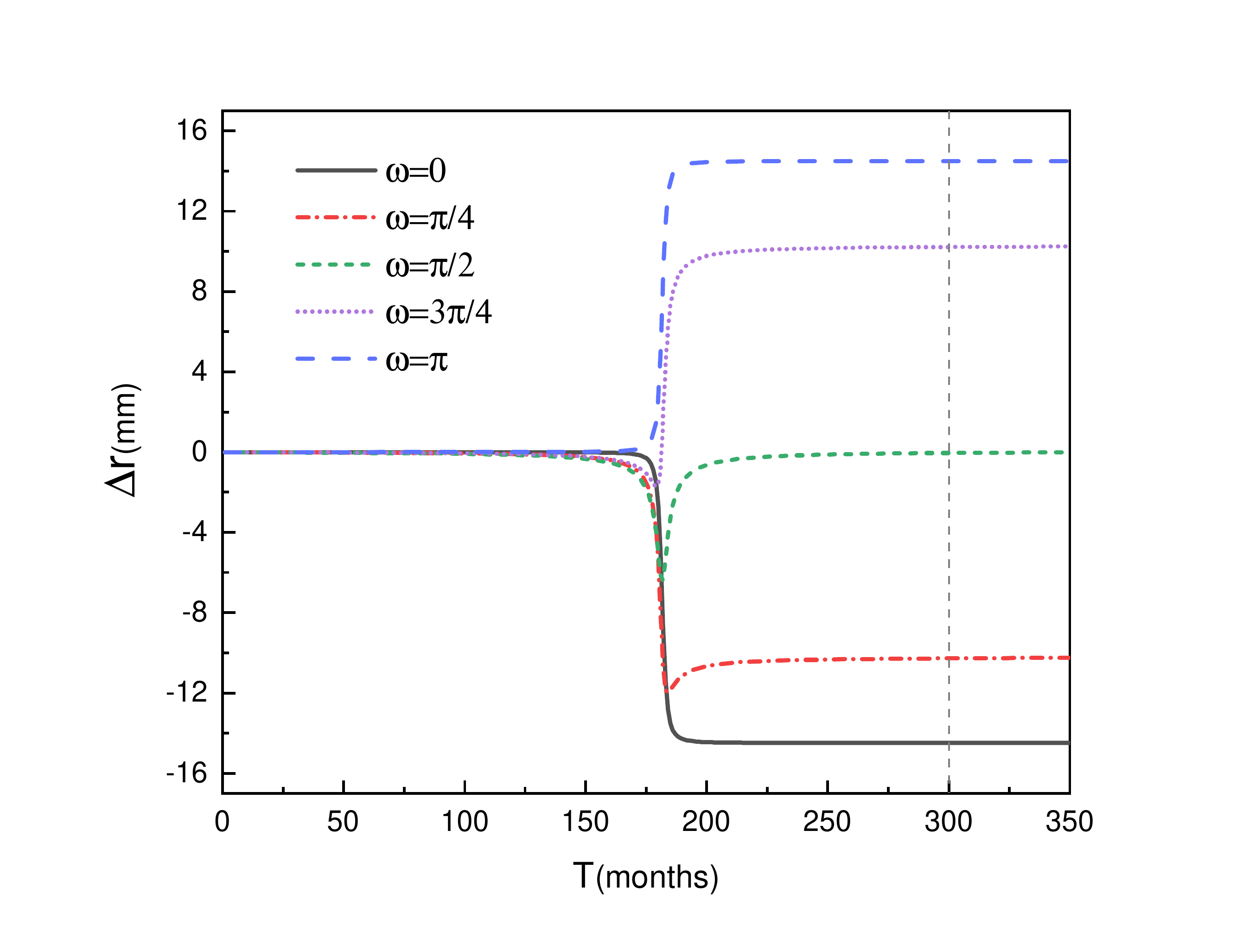}
	\includegraphics[scale=0.3]{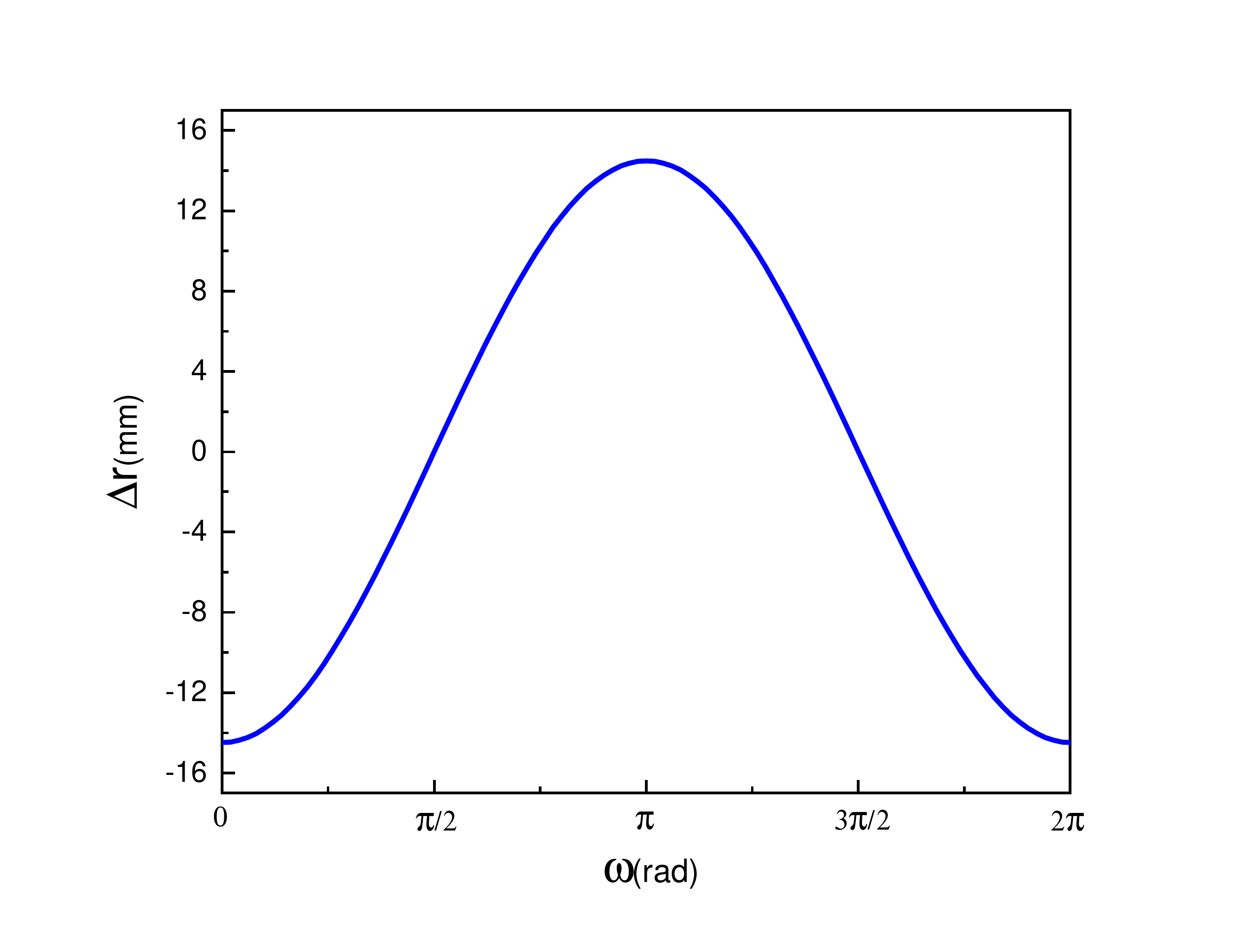}
	\caption{Left panel: the evolution of $\Delta r$ with time when $\varphi=0$ under different $\omega$ and other parameters fixed. Right panel:
		the values of $\Delta r(T=300)$ at the dotted line in the figure on the left panel as a function of $\omega$, which corresponds to the dependence of $\Delta r$ on $\omega$ when $c(t)=100b$.}
	\label{PBH-7}
\end{figure*}
\begin{figure*}[htbp]
	\centering
	\includegraphics[scale=0.3]{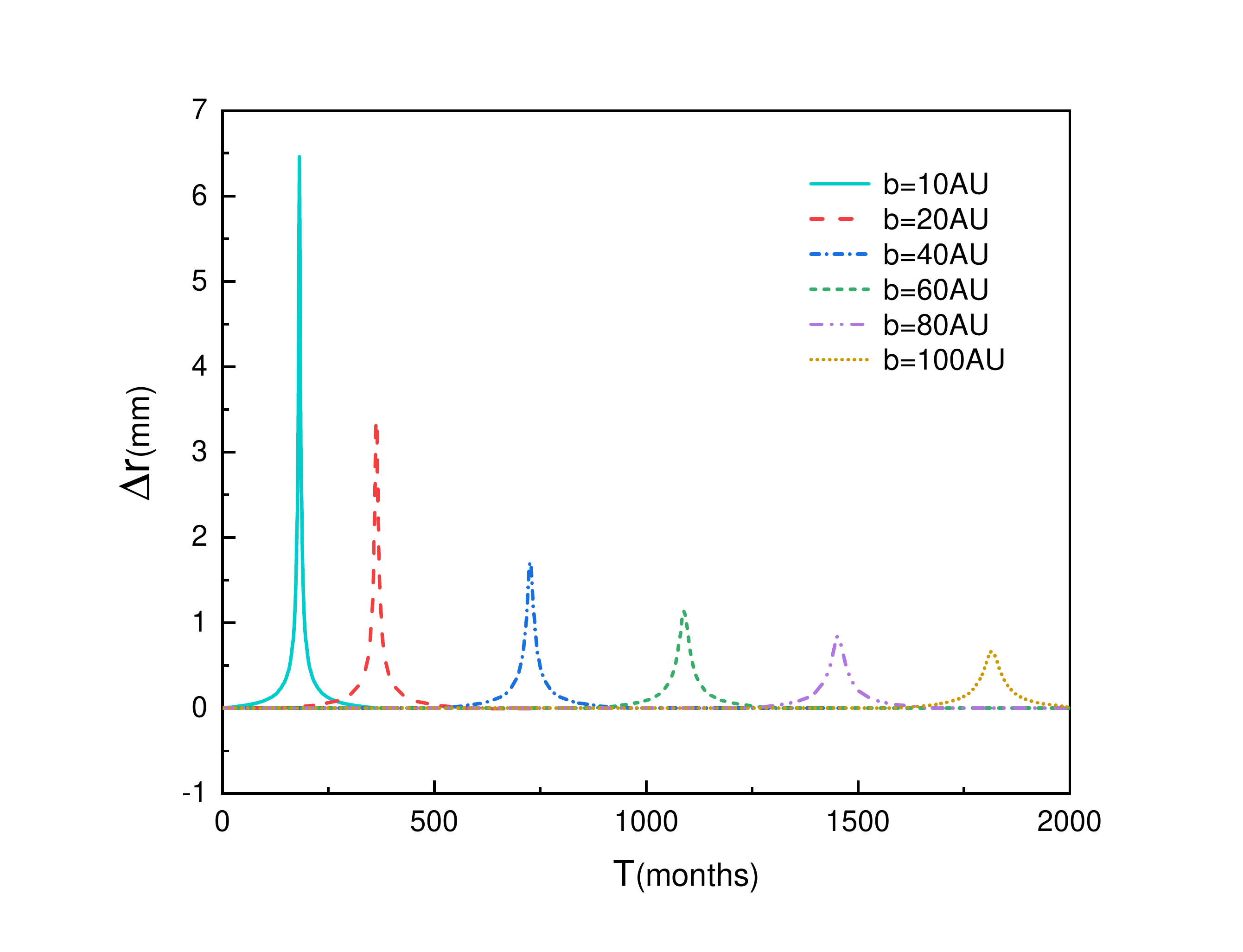}
	\includegraphics[scale=0.3]{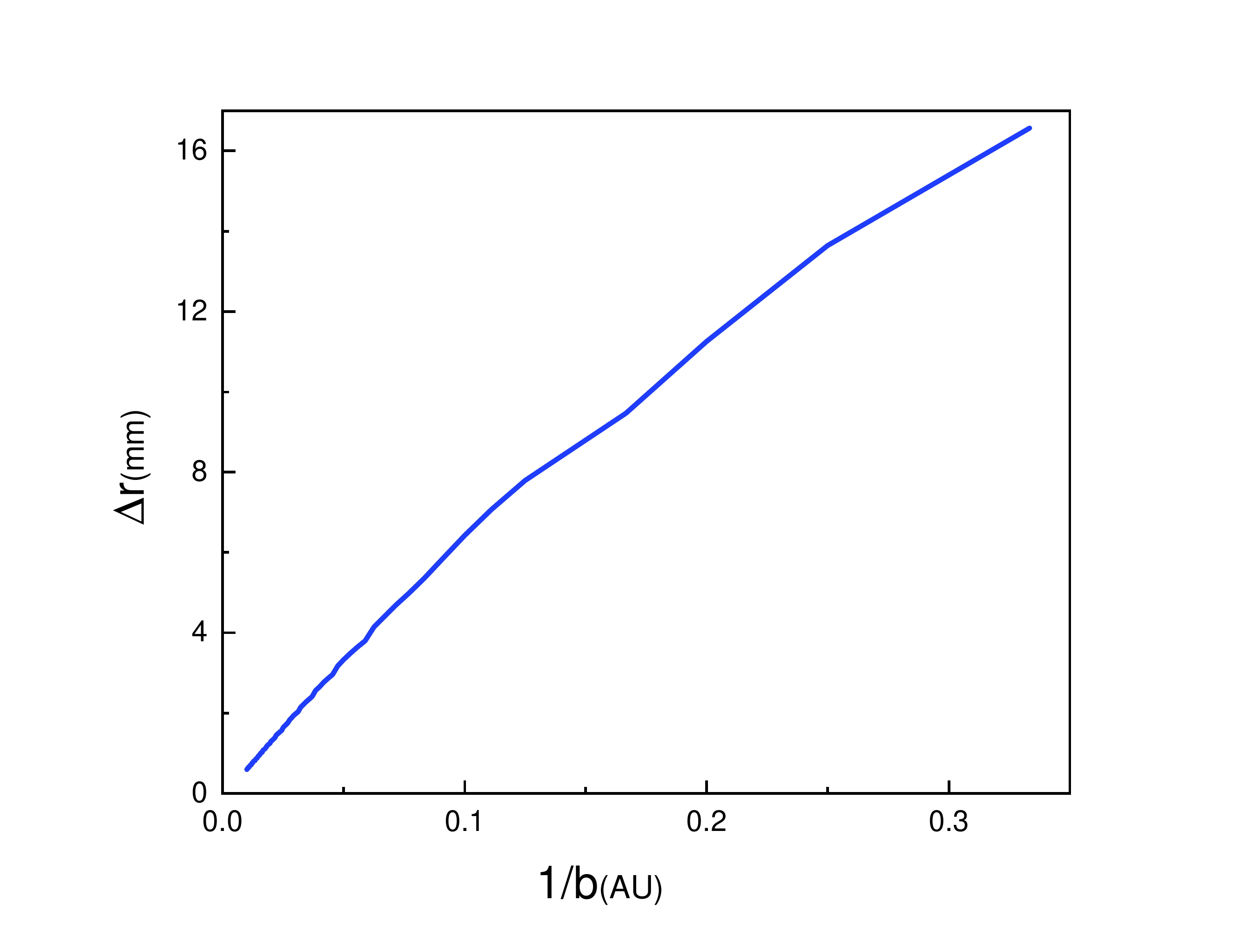}
	\caption{Left panel: the evolution of $\Delta r$ with time when $\varphi=\pi$ under different $b$ and other parameters fixed. Right panel: the peak value of $\Delta r(T)$ in the figure on the left panel as a function of $b$, which corresponds to the dependence of $\Delta r$ on $b$ when $c(t)=0$.}
	\label{PBH-8}
\end{figure*}
\begin{figure*}[htbp]
	\centering
	\includegraphics[scale=0.3]{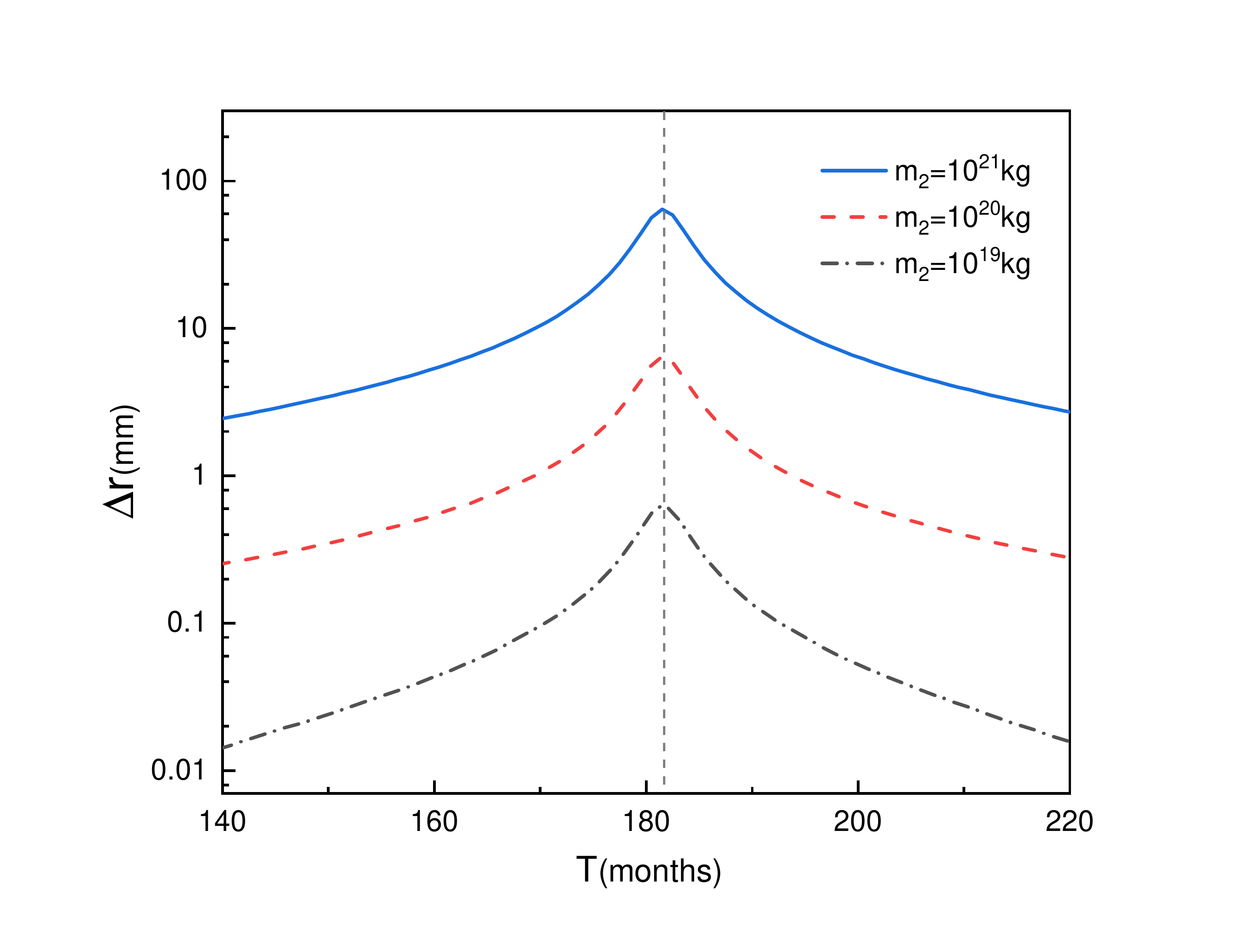}
	\includegraphics[scale=0.3]{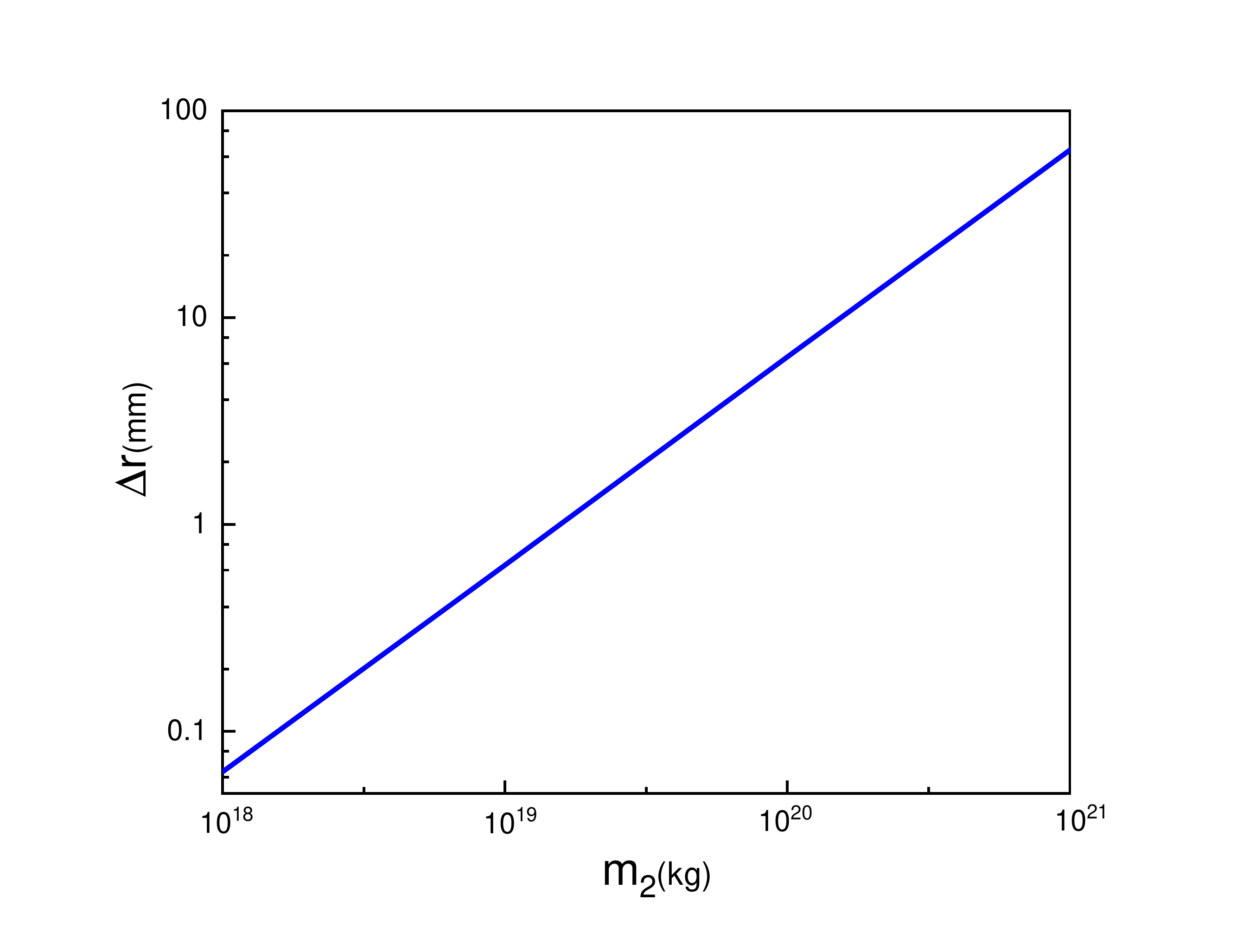}
	\caption{Left panel: the evolution of $\Delta r$ with time when $\varphi=\pi$ under different $m_{2}$ and other parameters fixed. Right panel: 
	the peak value of $\Delta r(T)$ (i.e., the values of $\Delta r$ at the dotted line in the figure on the left panel) as a function of $m_{2}$, which corresponds to the dependence of $\Delta r$ on $m_{2}$ when $c(t)=0$.}
	\label{PBH-9}
\end{figure*}
In what follows, let us turn to investigate the effects of the perturbing force on the Earth-Moon binary system when the observation angle $\varphi=0$ under different azimuths, distances and masses of the PBH, which is equivalent to vary a parameter in $i$, $\Omega$, $\omega$, $b$, $m_{2}$ and keep other parameters unchanged. The details are  shown in the left panels of Figs. \ref{PBH-5}-\ref{PBH-9}.
In addition, the dependence of $\Delta r$ on $i$, $\Omega$, $\omega$, $b$, $m_{2}$ are presented in the right panels of Figs. \ref{PBH-5}-\ref{PBH-9}, respectively.

From the left panel in Fig. \ref{PBH-5} and Fig. \ref{PBH-6}, we can see that simply changing $i$ has little effects on $\Delta r$, while changing $\Omega$ (with other parameters fixed) has larger effects on $\Delta r$. 
And $\Delta r$ can always return to zero in the end, regardless the values of $i$ and $\Omega$. This implies that the perturbing force of PBH has no long-term effects on the elliptical orbit of the Moon.
From the right panels in Fig. \ref{PBH-5} and Fig. \ref{PBH-6}, we can see that $\Delta r$ acts like a sinusoidal function of $i$ and $\Omega$ when $c(t)=0$, which is consistent with the calculated results in Ref. \cite{du2022probing}.

Fig. \ref{PBH-7}, however, presents another situation. It can be seen that $\Delta r$ cannot return to zero for any $\omega\neq  \frac{\pi }{2}$ or $\frac{3\pi }{2}$, which means that the perturbing force of PBH will have long-term effects on the elliptical orbit of the Moon. Again $\Delta r$ acts like a trigonometric function of $\omega$ as shown in the right panel in Fig. \ref{PBH-7}.

The left panel in Fig. \ref{PBH-8} shows that the widths of the peaks of the $\Delta r(T)$ curve are almost equal with different $b$, which means that the interaction time of the perturbing force of the PBH on the Earth-Moon binary system is almost the same, no matter how far the PBH is from the Earth.
In addition, the peak value of $\Delta r$ is almost inversely proportional to $b$, as clearly shown in the right panel of Fig. \ref{PBH-8}. This means that the farther the PBH from the Earth-Moon binary system is, the smaller the offset of the Moon's elliptical orbit will be.

Finally, let us turn to Fig. \ref{PBH-9}, which shows that $\Delta r$ is proportional to $m_{2}$, and is consistent with the results in Refs. \cite{seto2007searching,kashiyama2012enhanced}. Therefore, the greater the mass of the PBH is, the greater the offset of the Moon's elliptical orbit will be.
Roughly speaking, the heavier the PBHs are, the more possibilities to detect them are. However, PBHs with mass greater than the lunar mass ($\sim 10^{23}$kg) will make our perturbative calculations breakdown.

\maketitle
\newpage
\vspace*{0.2cm}

\subsection{When the initial position is non-coplanar}
We now discuss the influence of the perturbing force of the PBH on the Earth-Moon binary system, when the trajectory of the PBH is non-coplanar to the elliptical orbit of the Moon. 
Throughout the calculations of this subsection, we set the following initial condition: $i=\frac{\pi }{4}$, $\Omega =\frac{\pi }{4}$, $\omega =\frac{\pi }{4}$, $b=10$AU and $m_{2} =10^{20}$kg.

\begin{figure*}[htbp]
	\centering
	\includegraphics[scale=0.3]{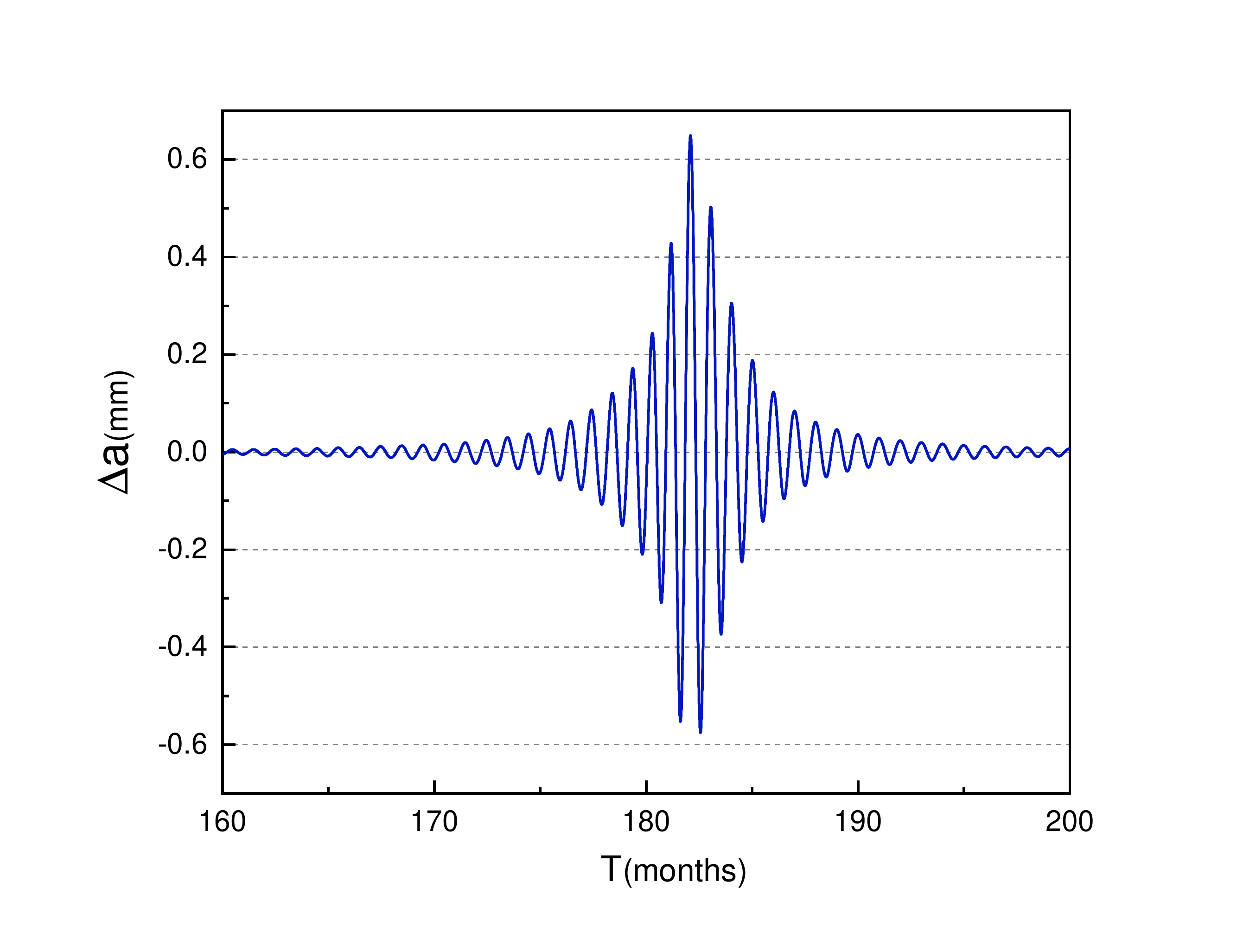}
	\includegraphics[scale=0.3]{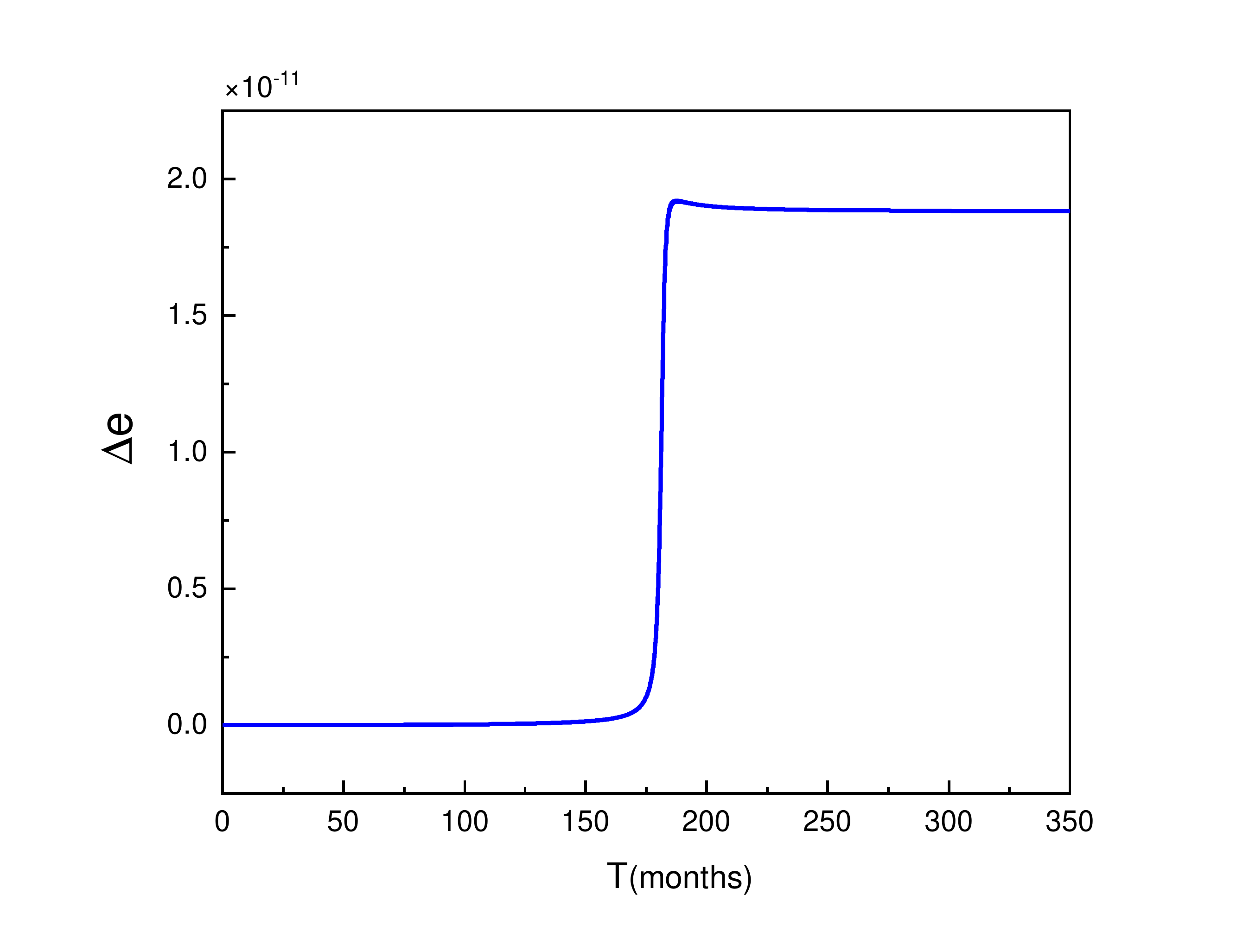}
	\includegraphics[scale=0.3]{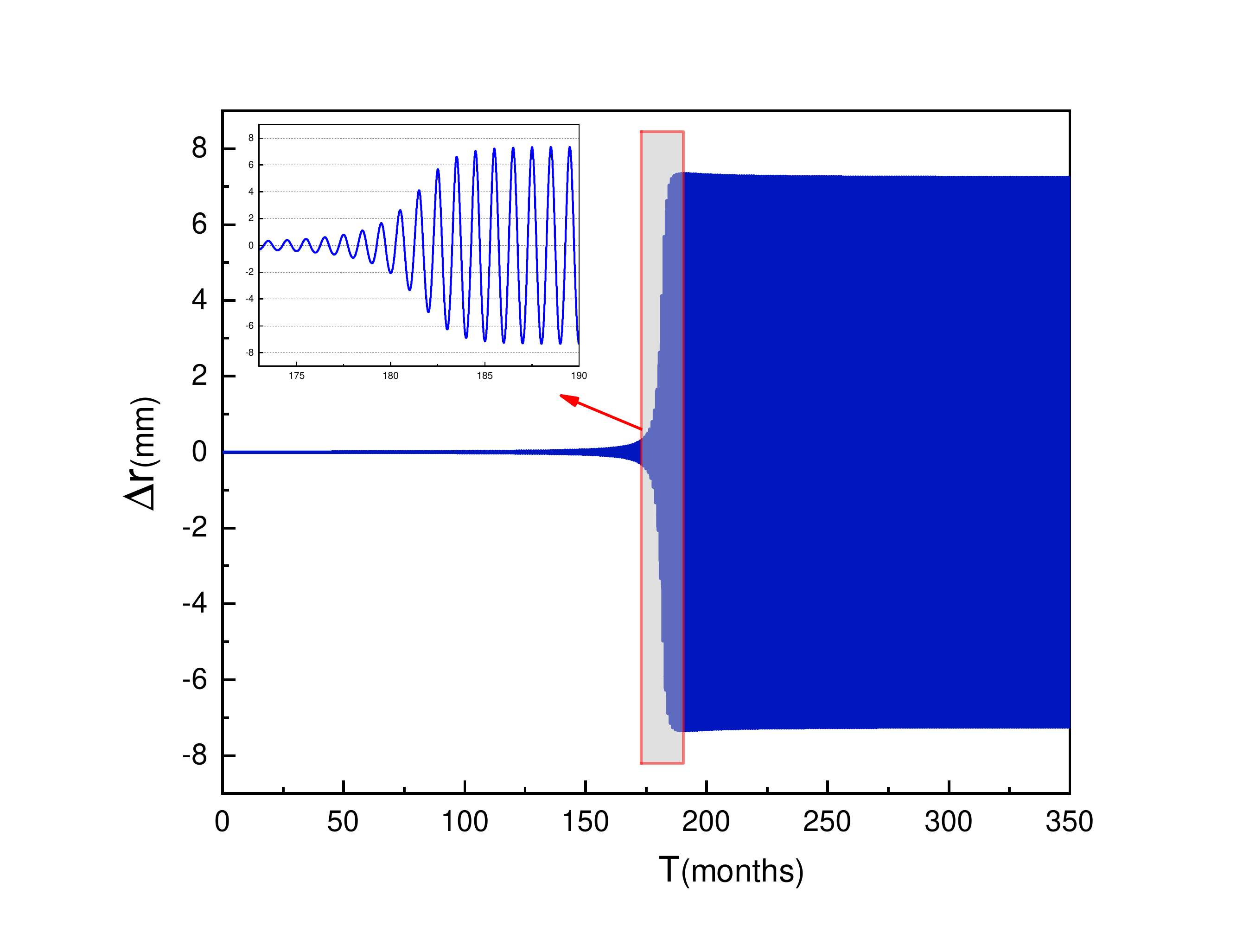}
	\caption{Evolution of $\Delta a$, $\Delta e$, $\Delta r$ with time.}
	\label{PBH-10}
\end{figure*}
Evolutions of $\Delta a$, $\Delta e$, $\Delta r$ with time are shown in Fig. \ref{PBH-10}. One can see that, after the PBH passes through, the semi-major axis of the orbit remains unchanged, while the eccentricity of the orbit increases which leads to the distance between the Earth and the Moon changing. This means that the perturbing force of PBH has long-term effects on the Earth-Moon binary system.

Comparing Fig. \ref{PBH-3} with Fig. \ref{PBH-10}, we find that $\Delta r$ is mainly determined by $\Delta e$.
This result can be explained by analyzing the relationship between $\Delta a$,  $\Delta e$ and  $\Delta r$ via Eq. \eqref{3}. 
According to Eq. \eqref{3}, $\Delta r$ is obtained by
\begin{equation}
	\label{30}
	\Delta r=\frac{1-e^{2}}{1+e \cos f} \Delta a+a\left[\frac{-2 e(1+e \cos f)-\left(1-e^{2}\right) \cos f}{(1+e \cos f)^{2}}\right] \Delta e,
\end{equation}
where $a \approx3.847\times10^{8}  $m and $e\approx0.055$ are the initial values of the semi-major axis and eccentricity of the Moon's elliptical orbit, respectively.
 When the observation point is at perigee, namely, $\varphi =0$, we have
\bqn
\label{31}	&\frac{1-e^{2}}{1+e \cos f}\approx 1,\\
\label{32}	&\frac{-2 e(1+e \cos f)-\left(1-e^{2}\right) \cos f}{(1+e \cos f)^{2}}\approx -1.
\eqn
Then
\begin{equation}
	\label{33}
	\Delta r\approx \Delta a-a \Delta e.
\end{equation}
 Combining Fig. \ref{PBH-10}, one can be seen that the magnitude of $\Delta a$ is much smaller than the term $a \Delta e$, therefore $\Delta r$ is mainly determined by $\Delta e$.

\begin{figure*}[htbp]
	\centering
	\includegraphics[scale=0.3]{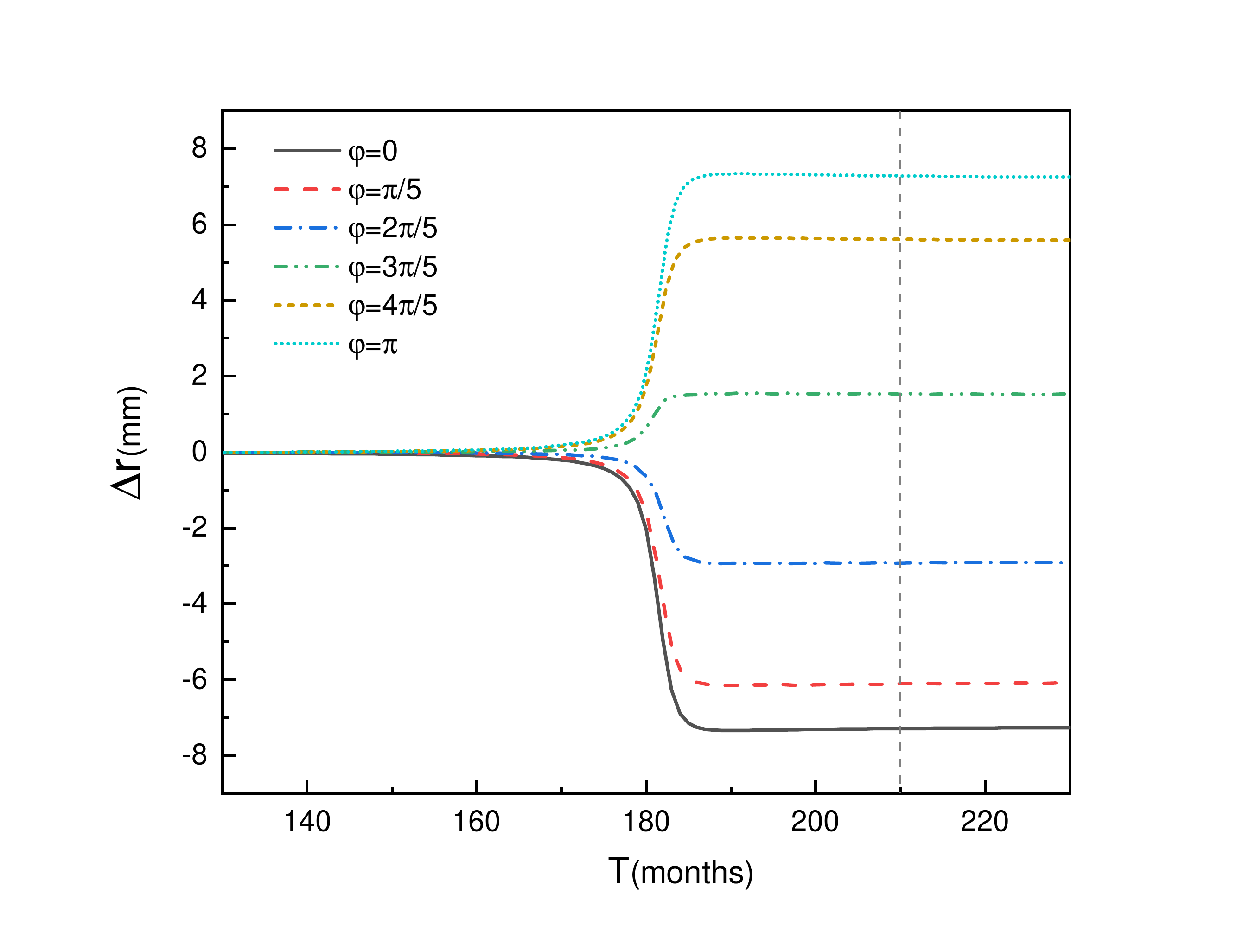}
	\includegraphics[scale=0.3]{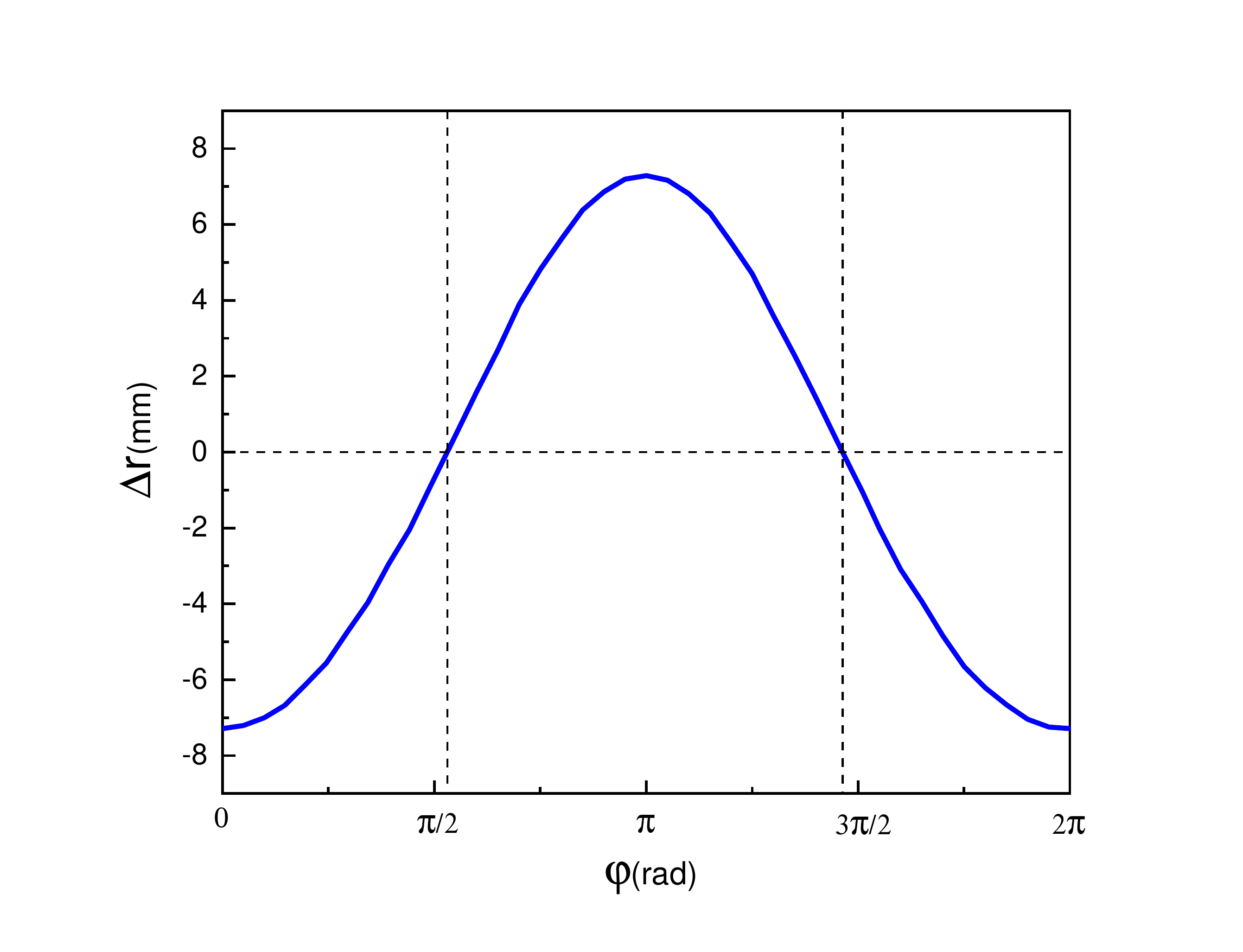}
	\caption{Left panel: evolution of $\Delta r$ with time under different fixed observation angles $\varphi$. Right panel: 
	the values of $\Delta r(T=210)$ at the dotted line in the figure on the left pane as a function of $\varphi$, which corresponds to the dependence of $\Delta r$ on $\varphi$ when $c(t)=100b$.}	
	\label{PBH-11}
\end{figure*}
The evolution of $\Delta r$ with time under different observation angles $\varphi$ and the dependence of $\Delta r$ on $\varphi$ are shown in Fig. \ref{PBH-11}. From the right plane of Fig. \ref{PBH-11}, one can see that the long-term effects of the perturbing force of PBH on the Earth-Moon distance are different under different observation angles. 
Combining the right plane of Fig. \ref{PBH-4} and Fig. \ref{PBH-11}, one can further see that the dependence of $\Delta r$ on $\varphi$ is the same when $i$, $\Omega $ and $\omega $ take different initial values. This means that $\Delta r$ always acts as a cosine function of $\varphi$, no matter where the orientation of the PBH is relative to the Earth-Moon binary system.

\begin{figure*}[htbp]
	\centering
	\includegraphics[scale=0.3]{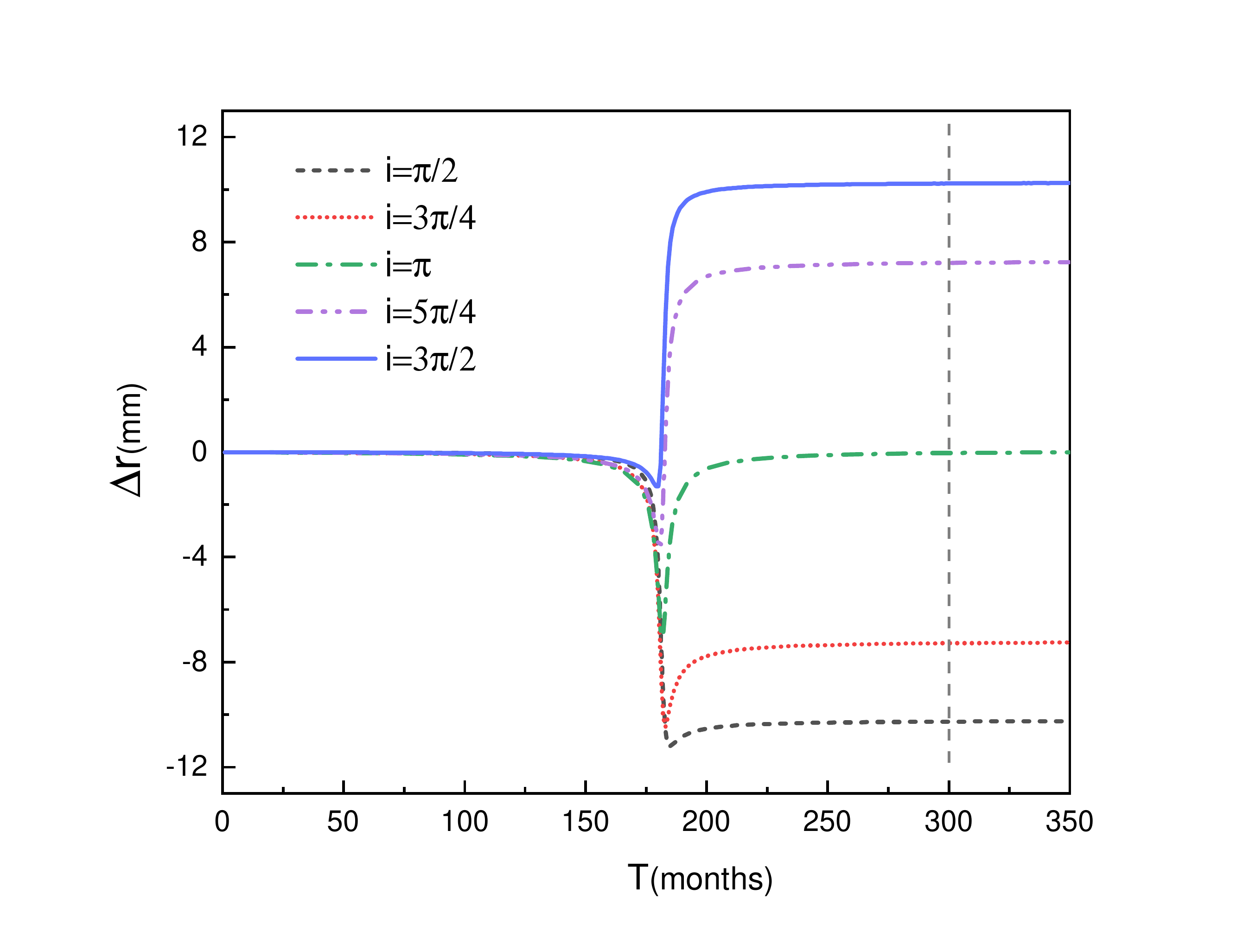}
	\includegraphics[scale=0.3]{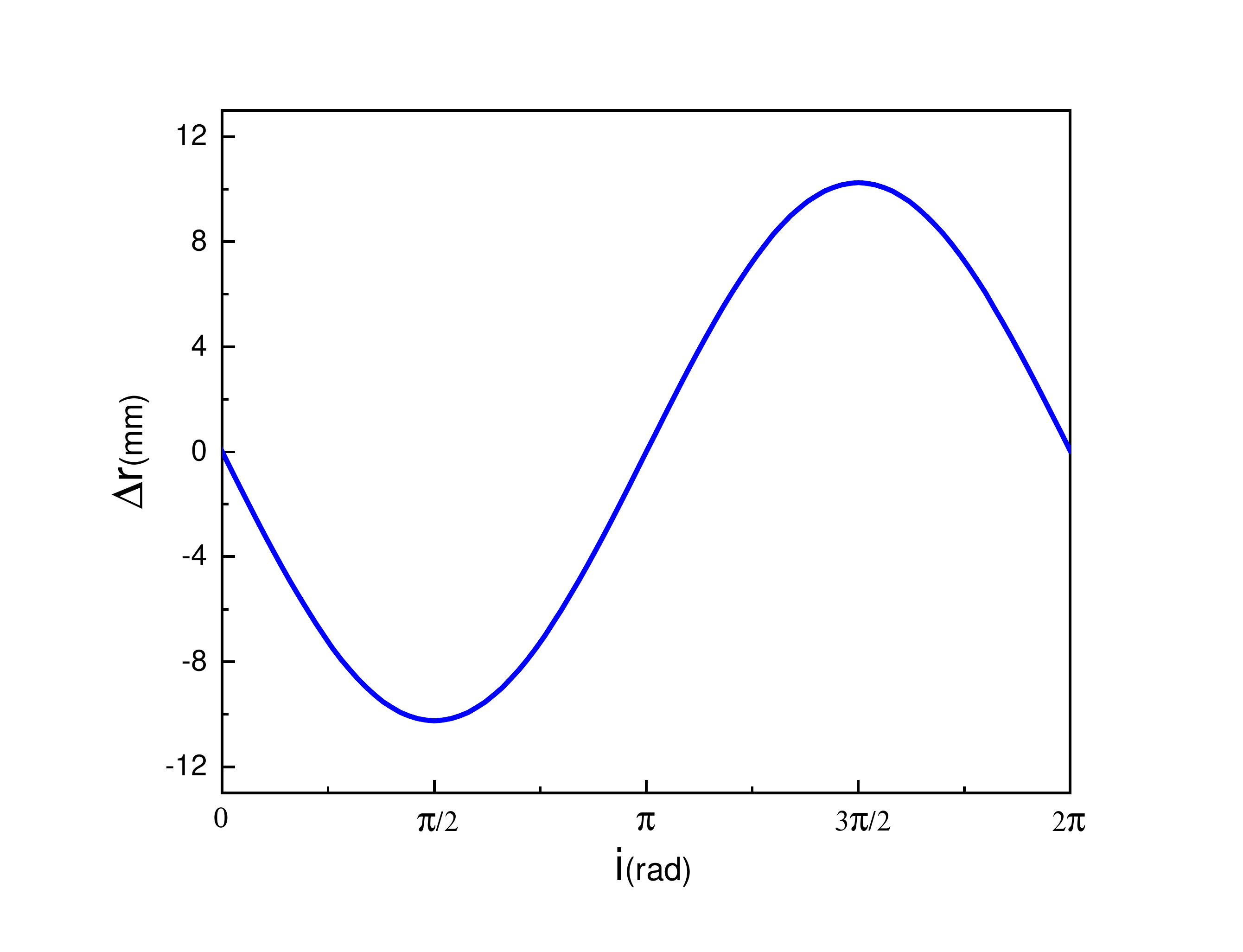}
	\caption{Left panel: evolution of $\Delta r$ with time when $\varphi=0$ under different $i$ and other parameters fixed. Right panel: 
		the values of $\Delta r(T=300)$ at the dotted line in the figure on the left panel as a function of $i$, which corresponds to the dependence of $\Delta r$ on $i$ when $c(t)=100b$.}
	\label{PBH-12}
\end{figure*}
\begin{figure*}[htbp]
	\centering
	\includegraphics[scale=0.3]{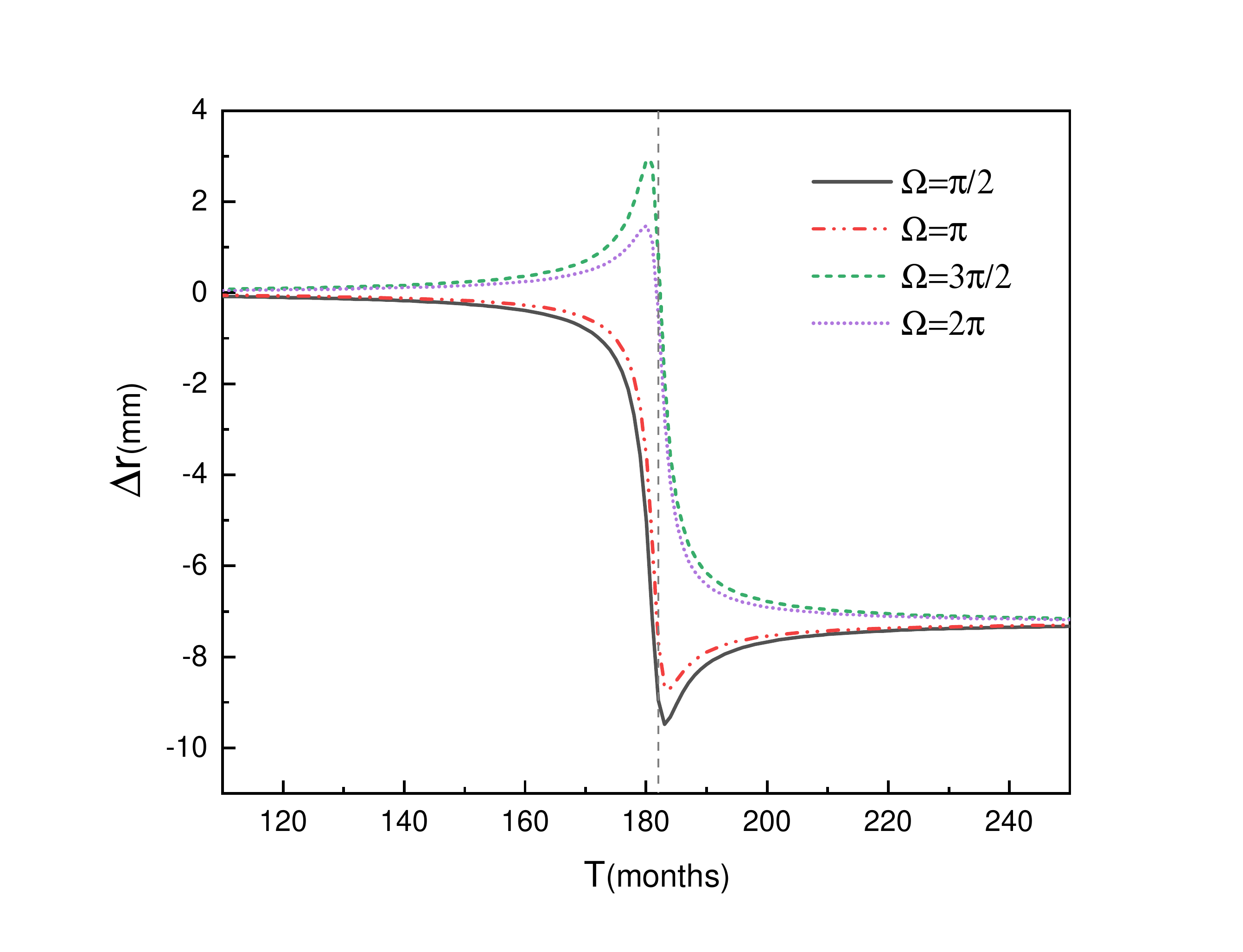}
	\includegraphics[scale=0.3]{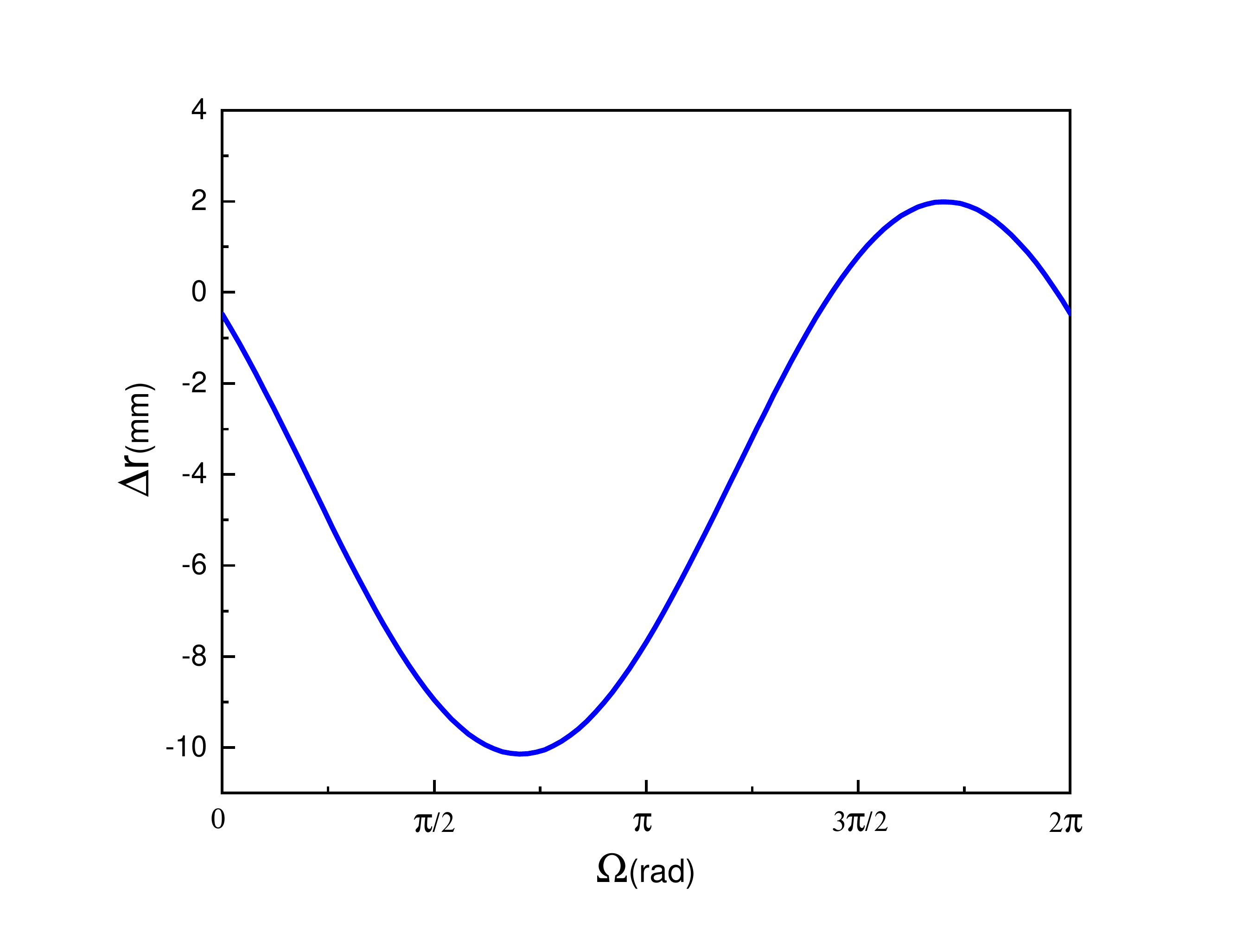}
	\caption{Left panel: the evolution of $\Delta r$ with time when $\varphi=0$ under different $\Omega$ and other parameters fixed. Right panel: 
		the values of $\Delta r(T=182)$ at the dotted line in the figure on the left panel as a function of $\Omega$, which corresponds to the dependence of $\Delta r$ on $\Omega$ when $c(t)=0$.}
	\label{PBH-13}
\end{figure*}
\begin{figure*}[htbp]
	\centering
	\includegraphics[scale=0.3]{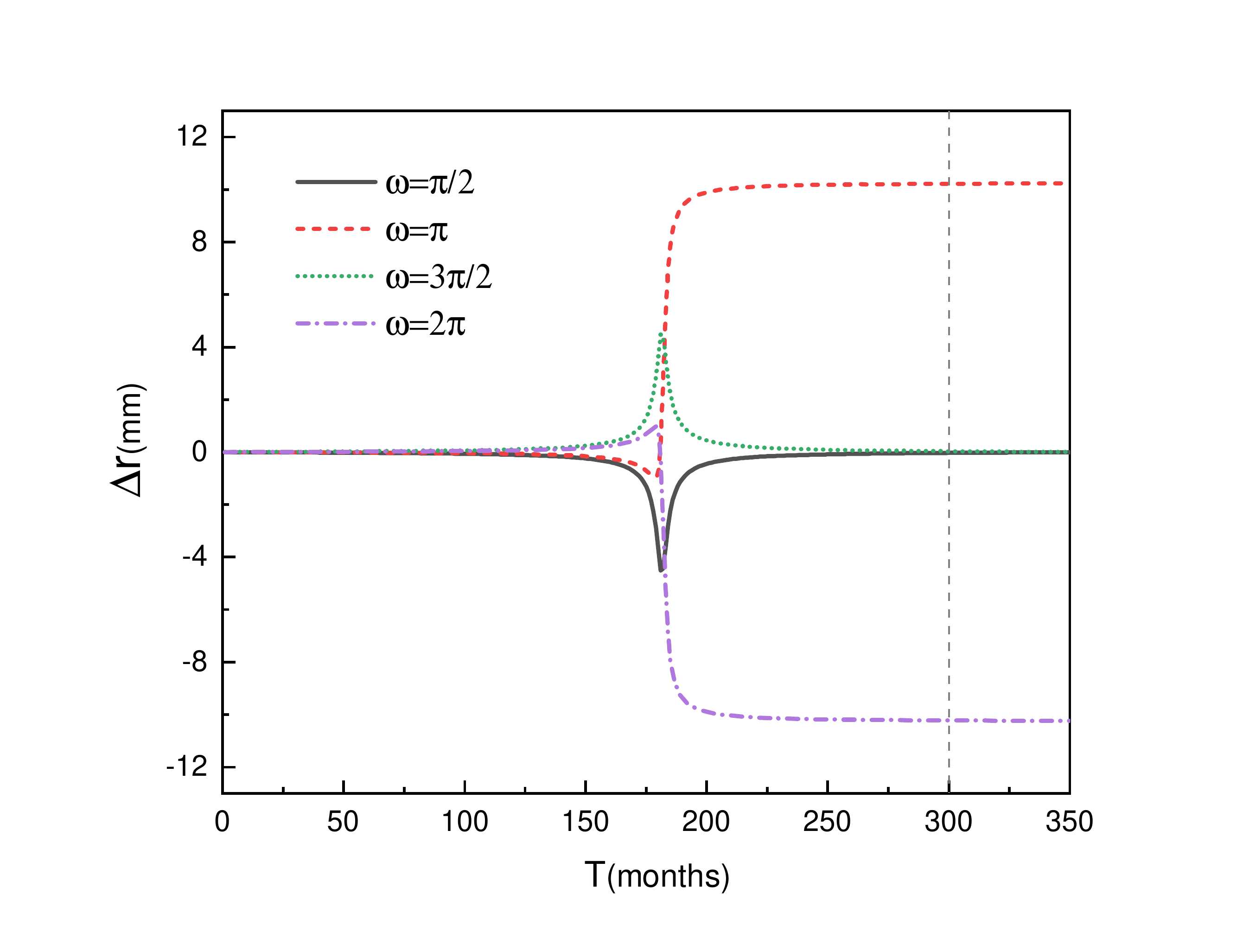}
	\includegraphics[scale=0.3]{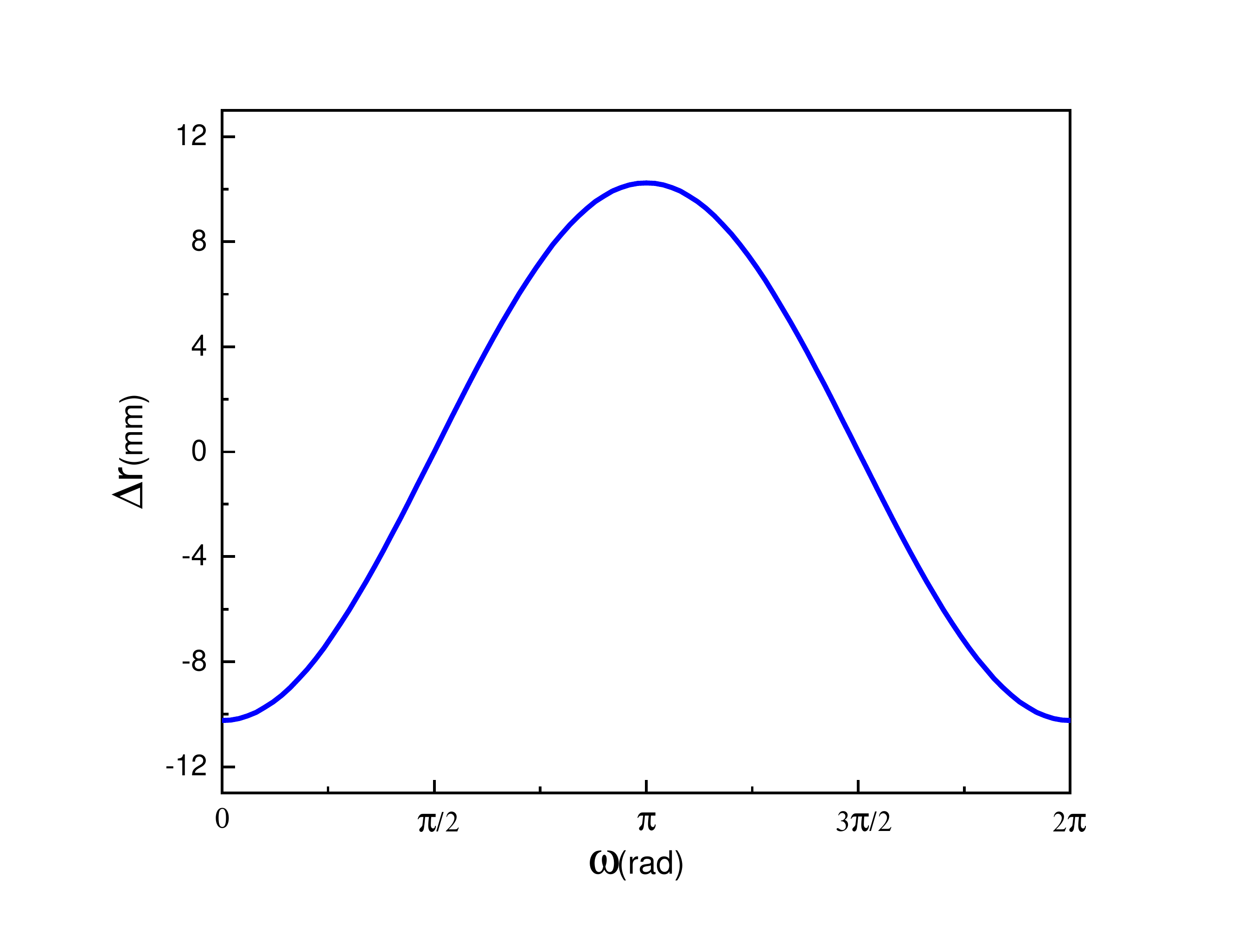}
	\caption{Left panel: the evolution of $\Delta r$ with time when $\varphi=0$ under different $\omega$ and other parameters fixed. Right panel: 
   the values of $\Delta r(T=300)$ at the dotted line in the figure on the left panel as a function of $\omega$, which corresponds to the dependence of $\Delta r$ on $\omega$ when $c(t)=100b$.}
	\label{PBH-14}
\end{figure*}
\begin{figure*}[htbp]
	\centering
	\includegraphics[scale=0.3]{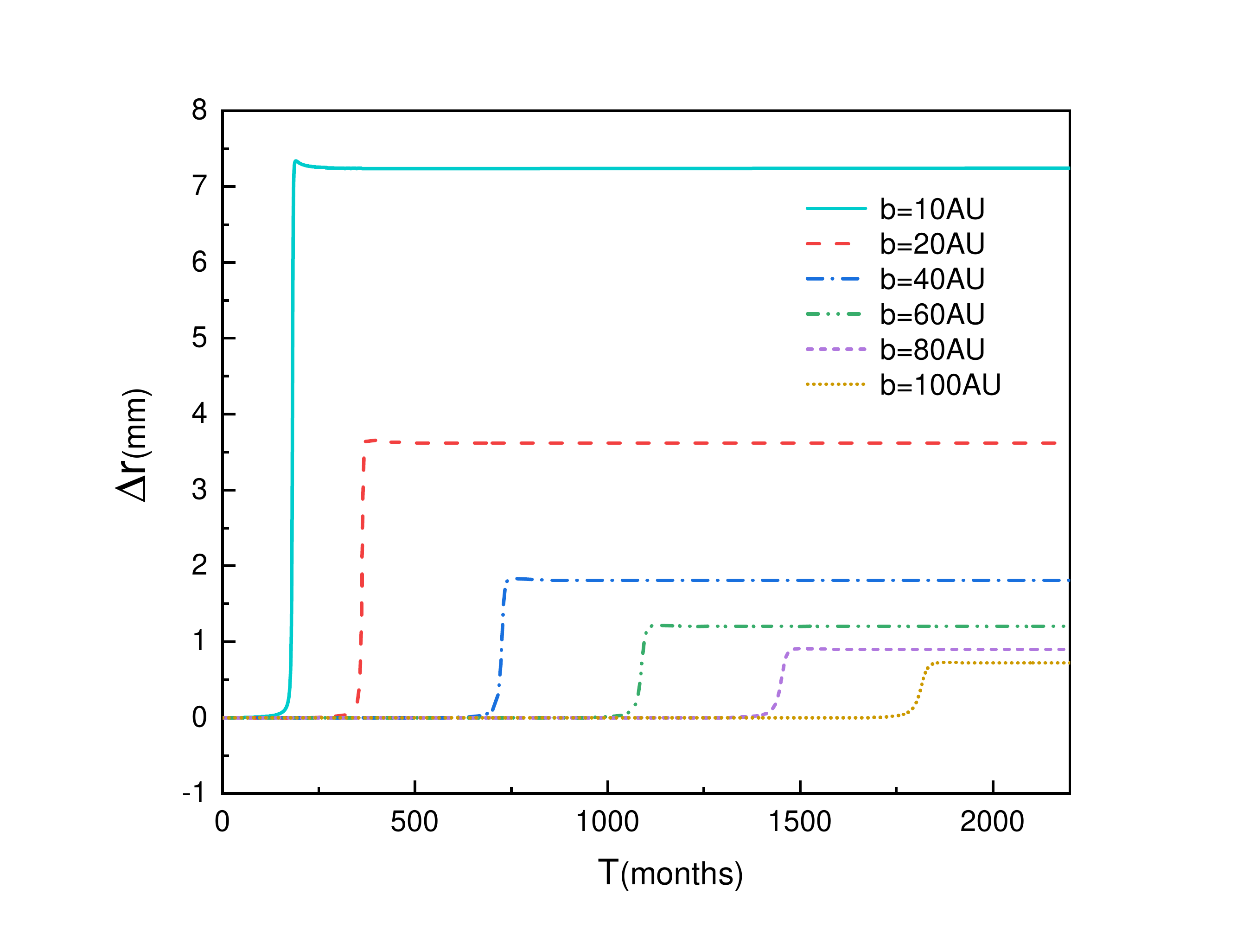}
	\includegraphics[scale=0.3]{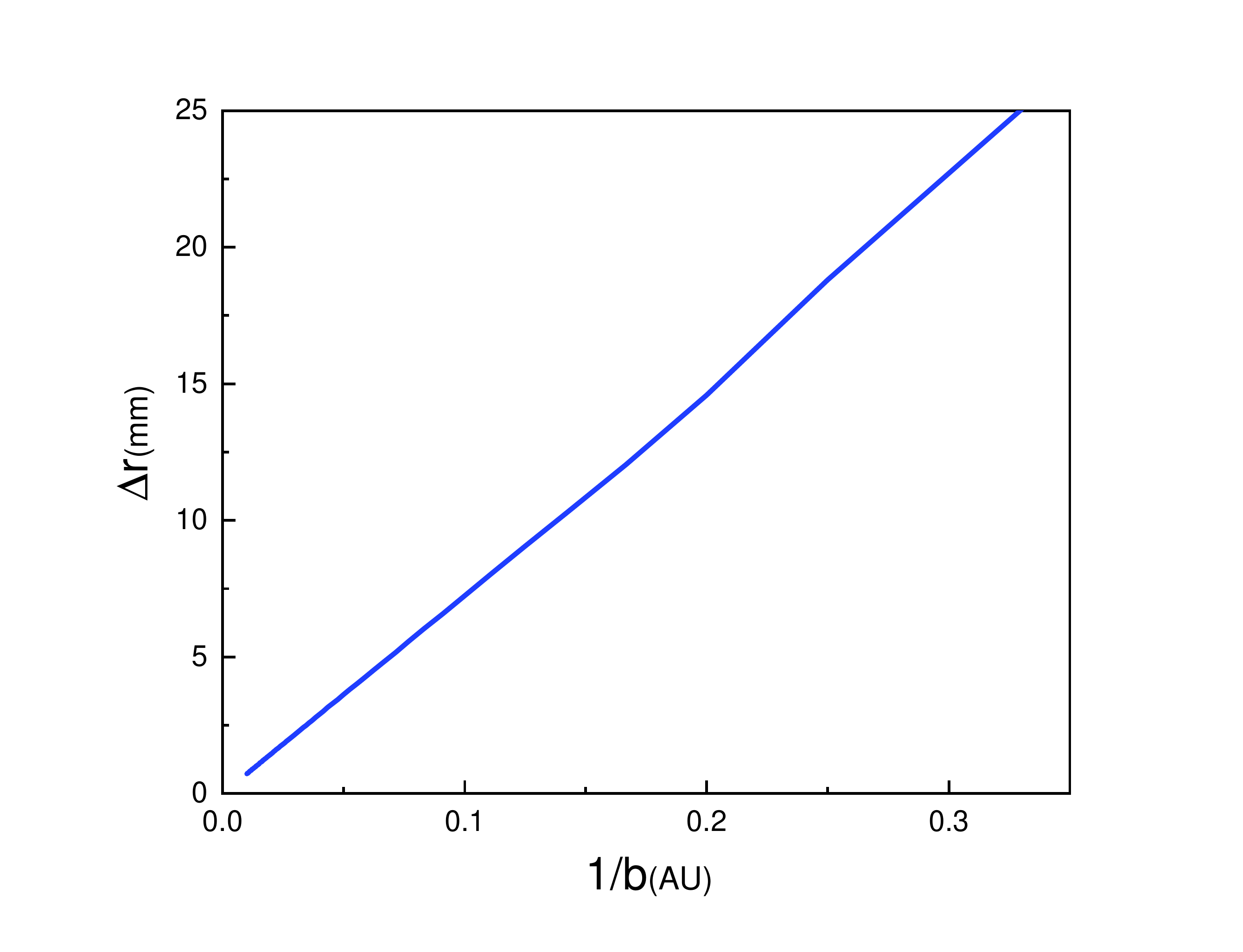}
	\caption{Left panel: evolution of $\Delta r$ with time when $\varphi=\pi$ under different $b$ and other parameters fixed. Right panel:  the dependence of $\Delta r$ on $\frac{1}{b} $ when $c(t)=100b$.}
	\label{PBH-15}
\end{figure*}
\begin{figure*}[htbp]
	\centering
	\includegraphics[scale=0.3]{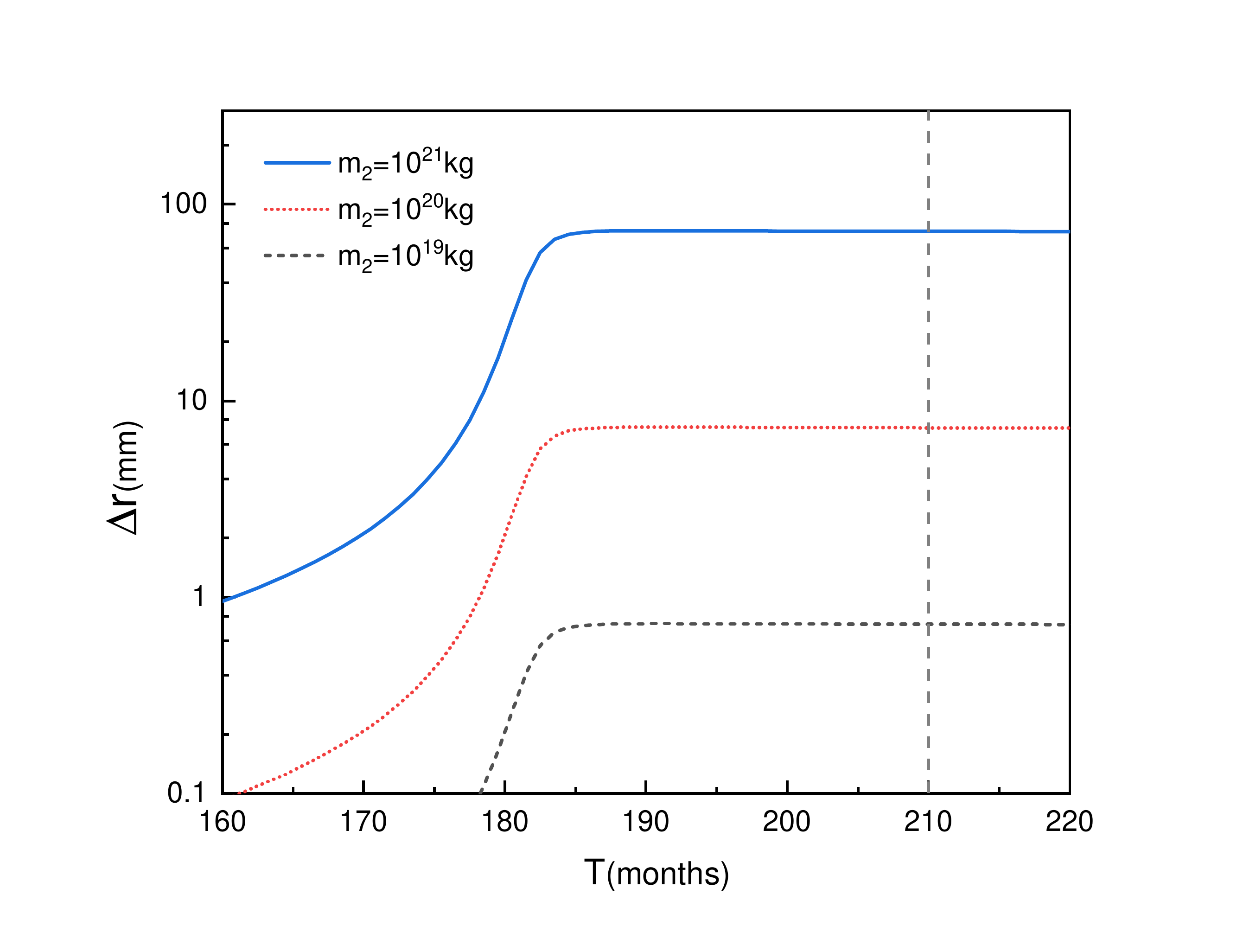}
	\includegraphics[scale=0.3]{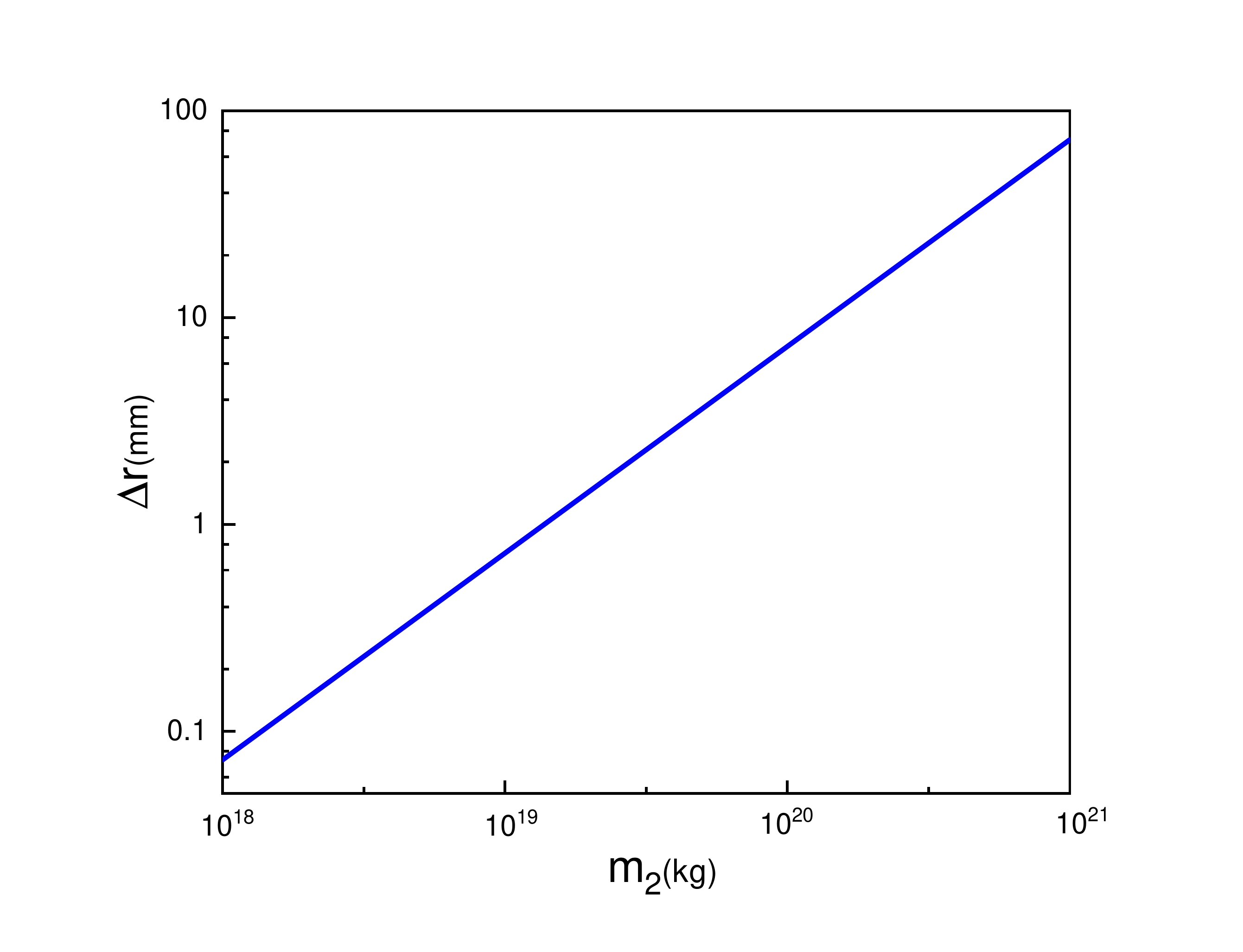}
	\caption{Left panel: evolution of $\Delta r$ with time when $\varphi=\pi$ under different $m_{2}$. Right panel:
		the values of $\Delta r(T=210)$ at the dotted line in the figure on the left panel as a function of $m_{2}$, which corresponds to the dependence of $\Delta r$ on $m_{2}$ when $c(t)=100b$.}
	\label{PBH-16}
\end{figure*}

The effects of the perturbing force of PBH on the Earth-Moon binary system under different azimuths, distances and masses of the PBH, which is equivalent to varying one of the parameters $i$, $\Omega$, $\omega$, $b$, $m_{2}$ and let other parameters fixed, are shown in the left panels of Figs. \ref{PBH-12}-\ref{PBH-16}, respectively.
The dependences of $\Delta r$ on $i$, $\Omega$, $\omega$, $b$, $m_{2}$, on the other hand, are shown in the right panels of Figs. \ref{PBH-12}-\ref{PBH-16}, respectively.

 From these figures, one can see that  $\Delta r$ is insensitive to $\Omega$ when $c(t)=100b$, comparing to $i$ and $\omega$, whose changes would have larger effects on $\Delta r$. In the coplanar case we see the same behavior.
Comparing the right panels of Fig. \ref{PBH-7} and Fig. \ref{PBH-14}, one can see that $\Delta r$ always acts as a cosine function of $\omega$, regardless of whether the PBH and the elliptical orbit of the Moon are coplanar or not.  
In addition, $\Delta r$ can always return to zero, as long as the value of $\omega$ is $\frac{\pi }{2}$ or $\frac{3\pi }{2}$, no matter what the values of $i$ and $\Omega $ are.
This point can be understood as follows. First, the elliptical orbit of Moon is perfectly symmetrical about the $Z$-axis when the value of $\omega$ is $\frac{\pi }{2}$ or $\frac{3\pi }{2}$. Additionally, the trajectory of the PBH is symmetrical about the $Z$-axis. This results in the work done by the perturbing force of PBH being equal to zero over the process. So the perturbing force of the PBH will not have long-term effects on the Earth-Moon binary system when the initial value of $\omega$ is $\frac{\pi }{2}$ or $\frac{3\pi }{2}$, no matter what the initial values of $i$ and $\Omega $ are. 
Combining Fig. \ref{PBH-8}, Fig. \ref{PBH-9}, Fig. \ref{PBH-15}, and Fig. \ref{PBH-16}, one can see that the dependence of $\Delta r$ on $b$ or $m_{2}$ is the same, regardless of whether the initial position of the PBH is coplanar with the Earth-Moon binary system or not.
 
 It is worth noting that in most cases, the values of $\Delta r$ are within the sensitivity of the current observations, such as the lunar laser ranging (LLR), whose ``normal point'' measurement can determine the Earth-Moon distance to as precision as $∼1$ mm \cite{murphy2013lunar}. That is to say, taking $b=10$ AU and  $c(t)=100b$ as an example,  signals for PBHs with mass larger than $10^{19}$kg should be able to detect by the ``normal point'' measurement.

\section{Conclusion}
In summary, we give a new proposal which can be used to detect sublunar-mass PBHs. In addition, by treating PBH as a perturbative term, we develop a framework to calculate the orbits of a generic binary system such as the Earth-Moon binary system. In order for the perturbative calculations trustable, one needs to assume that the PBH is far away from the Earth-Moon binary (far greater than $1$ AU), the mass of the PBH should be less than the Moon. These requirements constraint that the motivated PBH should have the mass less than $10^{22}$kg, a sublunar-mass PBH.  

Our numerical results show that the Earth-Moon distance is sensitive to the initial values of the system. Our discussion divides into two cases. One is for the case where the initial position of the PBH is coplanar to the the Earth-Moon binary system, the other is the opposite, i.e., the PBH is non-coplanar to the the Earth-Moon binary.  Both cases exhibits many similar behaviors. 
For example, in both cases, the Earth-Moon distance difference between with and without PBH $\Delta r$ always acts as a cosine function of $\varphi $, and  $\left| \Delta r\right| $ reaches its maximum value when $\varphi $ is equal to $0$ or $\pi $, which means that the greatest change in the distance between the Earth and the Moon can be observed by the ``normal point" measurement.
Similarly, in both cases, $\Delta r$ acts as a trigonometric function of $i, \omega$ and $\Omega$, and there always have long-term effects to the Earth-Moon orbits except for the case with $\omega = \frac{\pi}2, \frac{3\pi}2$, no matter what the initial values of other parameters are. Also, the long-term effects to the Earth-Moon orbits are insensitive to the initial value of $\Omega$, comparing to those of $i$ and $\omega$, whose initial values would have a significant impact on the long-term effects.
In addition, We observe the similar behavior of $\Delta r$ varying with the distance $b$ in both cases.  The peak value of $\Delta r$ gradually decreases as $b$ increases, which means that the further the PBH from the Earth-Moon binary system is, the smaller the offset of the Moon's elliptical orbit will be. Our results also show that $\Delta r$ is proportional to $m_{2}$, which is consistent with the calculated results in Refs. \cite{seto2007searching,kashiyama2012enhanced}. Therefore, the greater the mass of the PBH is, the greater the offset of the Moon's elliptical orbit will be.

Our results provide strong motivation for further work to develop the binary system as an accurate measurement tool for sublunar-mass PBHs.
It is of great interest to develop some data analysis pipelines to conduct PBHs searches with laser-ranging data, such that one can efficiently study the sublunar-mass PBHs,  and put new observational constraints on searching these sublunar-mass PBHs.

\section*{Acknowledgments}
This work is partially supported by the National Natural Science
Foundation of China with Grant No. 11975116.

\bibliographystyle{IEEEtran}
\bibliography{PBH}     %调用参考文献.bib

\end{document}